\documentclass{article}

\usepackage{arxiv}

\usepackage[utf8]{inputenc} 
\usepackage[T1]{fontenc}    
\usepackage{hyperref}       
\usepackage{url}            
\usepackage{booktabs}       
\usepackage{amsfonts}       
\usepackage{nicefrac}       
\usepackage{microtype}      

\usepackage{graphicx}

\usepackage{colortbl}		
\usepackage{multicol}        

\usepackage{amsbsy}
\usepackage{amsfonts}
\usepackage{amsmath}

\usepackage{amssymb,amsmath,mathrsfs,amsfonts,amsthm, amsbsy}
\usepackage{lineno,hyperref}
\usepackage{amsfonts}
\usepackage{amsmath}
\usepackage{xcolor}
\usepackage{booktabs} 
\modulolinenumbers[5]
\usepackage{comment}

\newtheorem{prop}{Proposition}

\newtheorem{defn}{Definition}

\newtheorem{pf_1}{Proof of Proposition \ref{prop::nssbss.sd}}
\newtheorem{pf_2}{Proof of Proposition \ref{prop::nssbss.jd}}
\newtheorem{pf_3}{Proof of Proposition \ref{prop::nssbss.jd2}}

\newcommand{\bs}[1]{\mathbf{#1}}
\newcommand{\x}{\bs x(\bs s)}
\newcommand{\z}{\bs z(\bs s)}
\newcommand{\s}{\bs s}
\newcommand{\A}{\bs A}
\newcommand{\W}{\bs W}
\newcommand{\U}{\bs U}
\newcommand{\V}{\bs V}
\newcommand{\T}{\bs T}
\newcommand{\M}{\bs M}
\def\Cov{\mathop{\mathrm{Cov}}\nolimits}

\def\E{\mathop{\mathrm{E}}\nolimits}
\newcommand{\dom}{\mathcal{S}}
\newcommand{\loc}{\mathcal{C}}
\newcommand{\R}{\mathbb{R}}

\title{Blind source separation for non-stationary 
random fields}

\date{} 					

\author{Christoph~Muehlmann \\
	Institute of Statistics \& Mathematical Methods in Economics \\
	Vienna University of Technology, Austria \\
	\texttt{christoph.muehlmann@tuwien.ac.at} \\
	\And
	Fran\c{c}ois~Bachoc\\
    Institut de Math\'{e}matiques de Toulouse \\
    Universit\'{e} Paul Sabatier, France \\
    \texttt{francois.bachoc@math.univ-toulouse.fr}
	\And
	Klaus~Nordhausen \\
	Department of Mathematics and Statistics \\
	University of Jyv\"askyl\"a, Finland \\
	\texttt{klaus.k.nordhausen@jyu.fi} \\}

\renewcommand{\headeright}{}
\renewcommand{\undertitle}{}

\begin{document}
\maketitle

\begin{abstract}
Regional data analysis is concerned with the analysis and modeling of measurements that are spatially separated by specifically accounting for typical features of such data. Namely, measurements in close proximity tend to be more similar than the ones further separated. This might hold also true for cross-dependencies when multivariate spatial data is considered. Often, scientists are interested in linear transformations of such data which are easy to interpret and might be used as dimension reduction. Recently, for that purpose spatial blind source separation (SBSS) was introduced which assumes that the observed data are formed by a linear mixture of uncorrelated, weakly stationary random fields. However, in practical applications, it is well-known that when the spatial domain increases in size the weak stationarity assumptions can be violated in the sense that the second order dependency is varying over the domain which leads to non-stationary analysis. In our work we extend the SBSS model to adjust for these stationarity violations, present three novel estimators and establish the identifiability and affine equivariance property of the unmixing matrix functionals defining these estimators. In an extensive simulation study, we investigate the performance of our estimators and also show their use in the analysis of a geochemical dataset which is derived from the GEMAS geochemical mapping project.
\end{abstract}

\section{Introduction} \label{sec:intro}

In spatial data analysis observations $x(\bs s_i)$, $i=1,\ldots,n$ are collected in a domain $\mathcal{S} \subset \mathbb{R}^d$ where $\bs s_i \in \mathcal{S}$ specifies the location of the observation $x(\bs s_i)$. In most applications $d=2$, which will be assumed in the following if not mentioned otherwise. It is meanwhile well-established that when analyzing spatial data the proximity of different observation locations has to be taken into account as observations located close to each other are expected to be more similar than observations further apart. The common way to consider this is via the covariance function
$$
C_x(\bs s_i, \bs s_j) = \E \left((x(\bs s_i)- \E (x(\bs s_i)))(x(\bs s_j)- \E (x(\bs s_j))\right).
$$ 

To make working with spatial data more tractable one assumes that the spatial observations are realizations of a weakly stationary random field which means one assumes that (i) $\E (x(\bs s_i)) = \mu$ for all $\bs s_i \in \mathcal{S}$ and that (ii) $C_x(\bs s_i, \bs s_j) = C_x(\bs s_i + \bs h, \bs s_j + \bs h)$, where $\bs h$ can be any shift with $\bs s_i,\bs s_j, \bs s_i+\bs h \ \mbox{and} \ \bs s_j + \bs h \in \mathcal{S}$. This assumptions state that the mean is constant over the domain and the covariance function is only a function of the difference between the sample locations but does not depend on the actual locations. This in turn means one can express the covariance function also as a one vector argument function, namely the difference $\bs h =  \s_i - \s_j$. If additionally the covariance function does only depend on the distance $h = \| \s_i - \s_j \|$ then its said to be isotropic. Usually, parametric covariance functions are specified and fitted to the data. One of the most popular parametric covariance function is the isotropic stationary M\'atern covariance function \cite{GuttorpGneiting2006}
\[
C(h;\sigma^2, \nu, \phi) = \frac{\sigma ^ 2}{2 ^ {\nu - 1} \Gamma (\nu)} \left( \frac{h}{\phi} \right) ^ \nu  K_\nu \left( \frac{h}{\phi} \right),
\]
where $K_\nu$ is the modified Bessel function of second kind, $\Gamma$ is the gamma function and $\sigma^2 > 0$, $\nu > 0$ and $\phi > 0$ are the variance, shape and range parameter respectively.

In many applications not only one variable is measured at each sample location but rather many, which yields multivariate spatial data where also cross-dependencies between the different variables have to be taken into account. Many suggestions and approaches for modeling the spatial cross-covariance functions for a $p$-variate random field $\bs x(\s)$,
$$
\bs C_{\bs x}(\bs s_i, \bs s_j) = E\left((\bs x(\bs s_i)-E(\bs x(\bs s_i)))(\bs x(\bs s_j)-E(\bs x(\bs s_j))^\top\right)
$$ 
are reviewed for example in \cite{GentonKleiber2015} where it is also pointed out that it is not that easy to create flexible and valid spatial  cross-covariance functions. One of the most popular approaches is the linear model of coreginonalization (LMC) \cite{GoulardVoltz1992,Wackernagel2003}, where the multivariate covariance function is formed by $r$ summands of $p \times p$ positive semi-definite coregionalization matrices $\bs T_k$ multiplied by univariate, parametric spatial correlation functions $\rho_k(h)$. Formally, the LMC is stated as

\[
	\bs C(h) = \sum^r_{k=1}  \bs T_k \rho_k(h).
\]

Another approach is followed by \cite{GneitingKleiberSchlather2010}, where the marginal and the cross-covariances are of the above M\'atern covariance form. The marginal covariance functions yield
\[
	C_{ii}(h;\sigma^2_{ii}, \nu_{ii}, \phi_{ii}) =\sigma^2_{ii} C(h; 1, \nu_{ii}, \phi_{ii}) \text{ for } i = 1, \dots, p, 
\] 
and the cross-covariances write as
\[
	C_{ij}(h;\rho_{ij}, \sigma_{ii}, \sigma_{jj}, \nu_{ij}, \phi_{ij}) = \rho_{ij} \sigma_{ii} \sigma_{jj} C(h; 1, \nu_{ij}, \phi_{ij}) \text{ for } i,j = 1, \dots, p, ~ i \neq j.
\] 
Conditions for the  shape, range, variance and correlation parameters $\nu_{ij}, \phi_{ij}, \sigma^2_{ii}$ and $\rho_{ij}$ for $i,j = 1, \dots, p$ which result in a valid multivariate cross-covariance function can be formulated, however, these conditions are rather involved and therefore the interested reader is referred to \cite{GneitingKleiberSchlather2010}. Similar as in the univariate case, the two above families of cross-covariance functions, and many others, make the assumption of weak stationarity and isotropy.

As the domains in modern applications are however often huge it is meanwhile commonly accepted that the weak stationarity assumption is convenient but not realistic. Stationarity seems rather justifiable on a local scale but not globally. Thus, recent years saw an increased interest in developing spatial methods which do not assume weak stationarity where the focus was mainly on univariate approaches. For example \cite{Sampson2010} reviews four different strategies to develop non-stationary covariance functions where the most popular approach seems to be based on spatial deformations. \cite{AnderesStein2011} focus on extending the M\'atern covariance for the non-stationary case by letting the shape, scale and variance parameters vary in the spatial domain. For the multivariate case \cite{VuZammitMangionCressie2021} point out that extensions of non-stationary cross-covariance functions are even more challenging to develop. \cite{GelfandSchmidtBanerjeeSirmans2004} extend the LMC to account for non-stationarity and \cite{KleiberNychka2012} extend the multivariate M\'atern model, both by introducing spatially varying parameters. \cite{VuZammitMangionCressie2021} on the other side extend the spatial deformation approach to the multivariate setting and also review some other approaches. In any way all the discussed approaches start with the selection of one or more cross-covariance functions which are then fitted to the data.

For multivariate spatial data recently \cite{NordhausenOjaFilzmoserReimann2015,BachocGentonNordhausenRuizGazenVirta2020}
suggested another approach, denoted as spatial blind source separation (SBSS). In SBSS the $p$-variate random field $\x$ is decomposed into $p$ uncorrelated / independent components which allows independent univariate modelling. However, SBSS also assumes weak stationarity. The goal of this work is to extend SBSS to the case of non-stationary spatial data which allows to discard the complex multivariate covariance modeling in favor of individual univariate modeling.

The structure of the paper is as follows. In Section~\ref{sec:model} we specify the exact considered spatial non-stationary blind source separation model. Three estimators for recovering the latent random fields are introduced in Section~\ref{sec:three_methods}, where the identifiability and affine equivariance properties of the underlying unmixing matrix functionals are studied. In an extensive simulation study we test the validity of our estimators in Section~\ref{sec:simulations} and illustrate their use on an environmental example in Section~\ref{sec:data_example}. Lastly, we conclude the paper in Section~\ref{sec:conclusion} and hint ideas for further research. The appendix contains the proofs of the stated propositions.

\section{A non-stationary spatial blind source separation model}\label{sec:model}
For the remainder of the paper we assume that the random field $\x$ at hand follows a spatial non-stationary (blind) source separation (SNSS) model which is defined as follows.
\begin{defn}[Spatial non-stationary source separation model]\label{def::nssbss}
A $p$-variate random field $\x$ defined on a $d$-dimensional spatial domain $\dom \subseteq \R ^ d$ follows a spatial non-stationary source separation model (SNSS) if it can be formulated as
\begin{equation}\label{eq::sbss_model}
 \x = \A \z + \bs b,
\end{equation}
where $\A$ is a deterministic invertible $p \times p$ mixing matrix, $\bs b$ is a $p$-variate deterministic location vector and $\z$ is a $p$-variate latent random field which fulfills the following assumptions
\begin{description}
 \item[(SNSS 1)] $\E (\z ) = \bs{0}$ for all $\s \in \dom$,
 \item[(SNSS 2)] $\Cov (\z ) = \E \left(\z \z ^ \top \right) = \bs{\Sigma}_\s$ where $\bs{\Sigma}_\s$ is a positive definite diagonal matrix for all $\s  \in \dom$ and
 \item[(SNSS 3)] $\Cov (\bs{z}(\bs{s}),  \bs{z}(\bs{s}') ) = \E (\bs{z}(\bs{s})  \bs{z}(\bs{s'}) ^ \top) = \bs{\Sigma}_{\bs{s} \bs{s'}},$ for all $\s \neq \s' \in \dom$ where $\bs{\Sigma}_{\bs{s} \bs{s'}}$ is a diagonal matrix depending on $\bs{s}$ and $\s'$.
\end{description}
\end{defn}

In practical considerations the random field $\x$ of Definition~\ref{def::nssbss} is observed on a set of $n$ deterministic sample locations $\loc = \{\s_1,\dots,\s_n\} \subset \dom$ which is a natural assumption for geostatistical applications. The domain $\dom$ can be thought of as a continuous version of the sample locations $\loc$ and can in principle be of any shape, but for convenience it is often a $d$-dimensional hyperrectangle, which so-to-speak covers $\loc$.  

Assumption (SNSS 1) states that the mean of each entry of the latent random field is a constant for the whole domain. In contrast, assumptions (SNSS 2) and (SNSS 3) allow the diagonal covariance as well as the diagonal spatial cross-covariance matrices to be dependent on the specific sample locations. In total, the observed random field is formed by uncorrelated latent random fields that are non-stationary in the sense that the second order dependencies are allowed to vary across the spatial domain. Often however the assumption of uncorrelated latent components is replaced by the stronger assumption of mutual independence. For general overviews of blind source separation (BSS) methods and their assumptions see for example \cite{comon2010handbook,NordhausenOja2018}. The SNSS model here can be seen as a spatial variant of the non-stationary time series model which is  for example considered in \cite{ChoiCichocki2000,ChoiCichocki2000b,ChoiCichockiBelouchrani2001,Nordhausen2014}.

 If (SNSS 2) and (SNSS 3) are forced to be stationary, i.e,  $\bs{\Sigma}_\s$ is constant and the diagonal matrix $\bs{\Sigma}_{\bs{s} \bs{s'}}$ carries stationary covariance functions on its diagonal elements, i.e. functions only of the difference vector $\bs h$ between $\s$ and $\s'$, then the model of Definition~\ref{def::nssbss} corresponds to the (stationary) SBSS model discussed in detail in \cite{NordhausenOjaFilzmoserReimann2015,BachocGentonNordhausenRuizGazenVirta2020}.

The main goal of SNSS is to recover the true latent random field $\z$ based on $\x$ alone. Thus, an unmixing matrix functional $\W = \W (\x)$ and a location functional $\T=\T(\x)$ are required such that $\z = \W \left(\x - \T \right)$. Note that assumptions (SNSS 1)-(SNSS 3) are not sufficient to make this a well-defined problem as the conditions do not fix the order, signs and scales of the latent components of $\z$. That is, let $\bs J = \bs P \bs S \bs D$ where $\bs P$ denotes a permutation matrix, $\bs S$ a sign-change matrix and $\bs D$ a diagonal matrix with positive diagonal values. Then, the pairs $(\bs A, \z)$ and $(\bs A \bs J^{-1}, \bs J \z)$ both lead to the same $\x$ and fulfill all requirements of Definition~\ref{def::nssbss}, hence, they are not distinguishable. This leads to the fact that recovering $\z$ is only possible up to order, signs and scale which is an ambiguity present in all BSS models and not considered a problem. For a detailed discussion about identifiability and general ambiguities in BSS models see for example \cite{BachocGentonNordhausenRuizGazenVirta2020,TongLiuSoonHuang1991,ErikssonKoivunen2004}.

Another requirement of unmixing matrix functionals is the affine equivariance property \cite{miettinen2015} which states that the same latent random field is recovered (up to order and sign) independent of the exact way of mixing. Let $\x$ be a random field and $\bs x^*(\s) = \bs B \x + \bs a$ its affine transformed version, where $\bs B$ is any invertible $p \times p$ matrix and $\bs a$ is any $p$-dimensional vector. For an affine equivariant unmixing matrix functional it holds that $\W (\bs B \x + \bs a) = \W (\x) \bs B ^{-1} = \A^{-1} \bs B ^{-1}$ up to order and sign of the row vectors. Multivariate statistical tools fulfilling this property belong to the more general invariant coordinate system (ICS) framework \cite{IlmonenOjaSerfling2012}.

The following definition formally states identifiability and the affine equivariance property of unmixing matrix functionals discussed before.

\begin{defn}[Unmixing matrix functional]\label{def::identifiability}
For a random field $\x$ following the SNSS model (Definition~\ref{def::nssbss}) a $p \times p$ matrix-valued functional $\W (\x)$ is an unmixing matrix functional if it satisfies:
\begin{description}
	\item[(Identifiability)]  $\W (\x) \A = \bs P \bs S \bs D$ for some permutation matrix $\bs P$, sign change matrix $\bs S$ and diagonal matrix with strictly positive diagonal elements $\bs D$.
	
	\item[(Affine equivariance)] $\W (\bs B \x + \bs a) = \bs P \bs S \W (\x) \bs B ^{-1}$ where $\bs B$ is an invertible $p \times p$ matrix, $\bs a$ is a $p$-dimensional vector, $\bs P$ is some permutation matrix and $\bs S$ is some sign change matrix. 
\end{description}
\end{defn}

In the subsequent section we introduce three unmixing matrix functionals that solve the above stated SNSS problem and investigate their identifiability and affine equivariance properties.

\section{Three SNSS methods}\label{sec:three_methods}

The goal of this section is to introduce three different unmixing matrix functionals $\W(\x)$ that can be used in conjunction with any location functional $\bs T(\x)$ to recover the latent random field $\z$ by  $\W(\x) \left(\x - \bs T(\x) \right)$. For $\bs T(\x)$ we simply use the expectation and in the following focus our discussion solely on $\W(\x)$. The key quantities for all three following unmixing matrix functionals are so-called local covariance matrices which are defined as

\begin{equation}\label{eq:lcov}
\M_{\dom,f}(\x) = \frac{1}{| \dom \cap \loc |} \sum_{\s_i,\s_j \in \dom \cap \loc} f(\s_i - \s_j) \E \Bigl[ [\bs x(\s_i) - \E (\bs x(\s_i))] [\bs x(\s_j)- \bs \E (\bs x(\s_j)) ]^\top \Bigr].
\end{equation}

Local covariance matrices were introduced in \cite{NordhausenOjaFilzmoserReimann2015} and refined in \cite{BachocGentonNordhausenRuizGazenVirta2020,muehlmann2020test} in the context of SBSS for the second order stationary case. Note that in Equation~\eqref{eq:lcov} we allow the considered spatial domain $\dom$ not to contain $\loc$ which is slightly different in comparison with the original definition, this will be useful when considering subdomains (see Section~\ref{sec:three_methods}). The matrices $\M_{\dom,f}(\x)$ compute a weighted average of the spatial covariances of all available pairs of coordinates $\dom \cap \loc$, where the weights are determined by the so-called spatial kernel function $f: \R ^d \rightarrow \R$. Three options are introduced in \cite{BachocGentonNordhausenRuizGazenVirta2020} as follows.

\begin{itemize}
	\item \textbf{Ball kernel:} $f_b(\s;r) = I(\| \s \| \leq r)$ where $r \geq 0$.
	\item \textbf{Ring kernel:} $f_r(\s;r_1,r_2) = I(r_1 < \| \s \| \leq r_2)$ where $r_1,r_2 \geq 0$ and $r_1 < r_2$.
	\item \textbf{Gauss kernel:} $f_g(\s;r) = \exp(-0.5 (\Phi^{-1}(0.95) \| \s \| / r)^2)$ where $r > 0$ and $\Phi^{-1}(0.95)$ is the $95\%$ quantile of the standard Normal distribution.
\end{itemize}

Here $I(\cdot)$ denotes the indicator function. All three kernel functions  above assume isotropic random fields as they only operate on the norm of $\s$. It is possible to define spatial kernel functions differently and account for possible anistropies present in the random fields, this is however beyond the scope of this paper.

For the special case of a ball kernel with parameter $r=0$, denoted as $f_0$, local covariance matrices reduce to the average covariance in $\dom$ where no spatial dependence is utilized. Formally
\[
	\M_{\dom,f_0}(\x) = \frac{1}{| \dom \cap \loc |} \sum_{\s \in \dom \cap \loc}  \E \Bigl[ [\bs x(\s) - \E (\bs x(\s))] [\bs x(\s)- \bs \E (\bs x(\s)) ]^\top \Bigr].
\]

Considering a finite sample, the estimation of the subsequently introduced mixing matrix functionals is carried out by replacing the population quantities from Equation~\eqref{eq:lcov} by their sample counterparts. Specifically, the corresponding sample version of Equation~\eqref{eq:lcov} is given by 
\begin{equation}\label{eq:sample_lcov}
\hat{\M}_{\dom,f}(\x) = \frac{1}{| \dom \cap \loc |} \sum_{\s_i,\s_j \in \dom \cap \loc} f(\s_i - \s_j)  (\bs x(\s_i) - \bar{\bs x})(\bs x(\s_j)- \bar{\bs x} )^\top ,
\end{equation}
where $\bar{\bs x} = n^{-1} \sum_{i=1}^n \bs x(\s_i)$, which also defines the sample version of $\M_{\dom,f_0}(\x)$. Additionally, we estimate the location functional $\bs T(\x)$ always by $\bar{\bs x}$.

For a random field $\x$ following the SNSS model (Definition~\ref{def::nssbss}) we observe that $\M_{\dom,f_0}(\z)$ as well as $\M_{\dom,f}(\z)$ yield diagonal matrices for all formerly discussed kernel function options which motivates the following three estimators.

\subsection{Simultaneous diagonalization of two average covariance matrices}

The first unmixing matrix functional is based on the simultaneous diagonalizaton (sd) of two average covariance matrices which is formalized in the following definition.

\begin{defn}[SNSS.sd functional]\label{def::nssbss.sd}
Consider a random field $\x$ following the SNSS model (Definition~\ref{def::nssbss}) and a partition of the spatial domain $\dom$ into $\dom_1,\dom_2$ where $\dom_1 \cap \dom_2 = \emptyset$. The SNSS.sd functional  $\W=\W(\x)$ is defined as the simultaneous diagonalizer satisfying
\[
\W \M_{\dom_1,f_0}(\x) \W^\top = \bs I_p \quad \mbox{and} \quad \W \M_{\dom_2,f_0}(\x) \W^\top = \bs D_{\dom_1 \dom_2},
\] 
where $\bs D_{\dom_1 \dom_2}$ is a diagonal matrix with decreasingly ordered diagonal elements.
\end{defn}

Given a sample, an unmixing matrix $\W$ can be found by solving the generalized eigenvalue-eigenvector problem, which always yields exact diagonalization of the former two matrices. Furthermore, the decreasing ordering of the diagonal elements of $\bs D_{\dom_1 \dom_2}$ comes without loss of generality as the order of the latent random field can be anyhow only recovered up to permutations. The following proposition gives a necessary condition for the identifiability of the the above unmixing matrix functional as well as the desired affine equivariance property.

\begin{prop}\label{prop::nssbss.sd}
The SNSS.sd functional seen in Definition~\ref{def::nssbss.sd} is 
\begin{description}
\item[(1)] identifiable as seen in Definition~\ref{def::identifiability} if and only if elements of the diagonal matrix $\M_{\dom_1,f_0}^{-1}(\z) \M_{\dom_2,f_0}(\z)$ are pairwise distinct,

\item[(2)] affine equivariant as seen in Definition~\ref{def::identifiability}.
\end{description}
\end{prop}

According to \cite[Result 1]{Nordhausen2014} to ensure identifiability there need to exist at least two locations $\s_1,\s_2 \in \loc$ for which the elements of the diagonal matrix $\bs{\Sigma}^{-1}_{\s_1} \bs{\Sigma}_{\s_2}$ are pairwise distinct (where $\bs{\Sigma}_{\s}$ refers to the covariance matrices from Definition~\ref{def::nssbss}). If the former holds then it is possible to find two disjoint sub-domains $\dom_1,\dom_2$ of $\dom$ in such a way that all elements of the diagonal matrix $\M_{\dom_1,f_0}^{-1}(\z) \M_{\dom_2,f_0}(\z)$ are pairwise distinct. Note that \cite[Result 1]{Nordhausen2014} is formulated for the times series non-stationary blind source separation model, the above outline is the natural extension of this statement to the spatial non-stationary case. However, in practical considerations the desired partition is unknown, therefore the a-priori choice of the partition of the domain is not trivial and greatly affects the performance of the method. This issue is addressed in the following extension of the former unmixing matrix functional.

\subsection{Joint diagonalization of more than two average covariance matrices}

In contrast to the former method the spatial domain is divided into more than two subdomains and the corresponding average covariance matrices are jointly diagonalized (jd) as follows.

\begin{defn}[SNSS.jd functional]\label{def::nssbss.jd}
Consider a random field $\x$ following the SNSS model (Definition~\ref{def::nssbss}). Standardize $\x$ by $\bs x^{st}(\s) = \M_{\dom,f_0}^{-1/2}(\x)(\x - \bs b)$ and partition the spatial domain $\dom$ into $\dom_1,\dots,\dom_K$ where $\dom_m \cap \dom_n = \emptyset$ for $m,n=1,\dots,K$ and $m \neq n$. Then, let $\U$ be the orthogonal $p \times p$ joint diagonalizer of the matrices $\M_{\dom_k,f_0}(\bs x^{st}(\bs s))$ for $k=1,\dots,K$, which maximizes
\[
\sum_{k=1}^K \| \text{diag}(\U \M_{\dom_k,f_0}(\bs x^{st}(\s)) \U^\top )\|^2_F.
\] 
Then, the SNSS.jd functional equals $\W (\x) = \U \M_{\dom,f_0}^{-1/2}(\x)$.
\end{defn}

In the above definition $\text{diag}(\cdot)$ is a diagonal matrix with the diagonal elements equalling the ones of the matrix-valued argument, and $\|\cdot\|_F$ denotes the Frobenius norm. $\U$ is denoted an orthogonal joint diagonalizer of the matrices $\M_{\dom_k,f_0}(\bs x^{st}(\s))$ for $k=1,\dots,K$ as maximizing the diagonal elements is equal to minimize the off-diagonal elements by the orthogonal invariance of the Frobenius norm. Note that for a finite sample, usually the sample versions of the matrices $\M_{\dom_k,f_0}(\bs x^{st}(\s))$ for $k=1,\dots,K$ given by Equation~\eqref{eq:sample_lcov} do not commute, hence, exact joint diagonalization is impossible. Therefore, algorithms that find an approximate joint diagonalizer are needed. We choose one such algorithm that relies on Givens rotations \cite{CardosoSouloumiac1996}, but many others are available, see for example \cite{IllnerMiettinenFuchsTaskinenNordhausenOjaTheis2015}.

The next proposition is concerned with identifiability as well as affine equivariance.

\begin{prop}\label{prop::nssbss.jd}
The SNSS.jd functional seen in Definition~\ref{def::nssbss.jd} is
\begin{description}
\item[(1)] identifiable iff for all pairs $i,j=1,\dots,p$ and $i \neq j$ there exists a $k \in \{1,\dots,K\}$ such that $(\M_{\dom,f_0}^{-1}(\z) \M_{\dom_k,f_0}(\z))_{ii} \neq (\M_{\dom,f_0}^{-1}(\z) \M_{\dom_k,f_0}(\z))_{jj}$,

\item[(2)] affine equivariant as seen in Definition~\ref{def::identifiability}.
\end{description}
\end{prop}

The condition for identifiability given in Proposition~\ref{prop::nssbss.jd} is more general than the one in Proposition~\ref{prop::nssbss.sd} as a finer partition of the domain is allowed. Therefore, the exact partition of the domain for the SNSS.jd method should have less influence on the performance as long as enough sub-partitions are considered. In practical applications it might be useful to simply overlay the spatial domain by a grid formed by equally sized squared shaped blocks which define the sub-division of $\dom$, a procedure that we investigate in more detail in the simulation study in Section~\ref{sec:simulations}. The advantage of less sensitivity on the exact domain sub-partition of the the SNSS.jd methods comes at the cost of giving up exact diagonalization from the SNSS.sd method, which introduces more computational complexity as joint diagonalization algorithms need to be applied. 

Both former methods have in common that only the spatial ordering of the points is taken into account but not the spatial dependencies between them when computing the unmixing matrix. A trivial example which would cause problems is the case when the matrices $\bs \Sigma_\s$ are the identity matrix for all $\s \in \dom$ but $\bs{\Sigma}_{\bs{s} \bs{s'}}$ is non-zero and spatial dependent. In that case the identifiabilty conditions of Propostions~\ref{def::nssbss.jd} and consequently the one of Propostion~\ref{def::nssbss.sd} do not hold and the two methods fail. In that case recovering the latent random field is still possible when considering second order spatial dependencies as suggested in the following approach.

\subsection{Joint diagonalization of more than two local covariance matrices}

The following SNSS.sjd divides the domain into at least two parts and jointly diagonalizes the corresponding local covariance matrices for a set of kernel functions, therefore, it utilizes second order spatial dependence (sjd). 

\begin{defn}[SNSS.sjd functional]\label{def::nssbss.jd2}
Consider a random field $\x$ following the SNSS model (Definition~\ref{def::nssbss}). Standardize $\x$ by $\bs x^{st}(\s) = \M_{\dom,f_0}^{-1/2}(\x)(\x - \bs b)$ and partition the spatial domain $\dom$ into $\dom_1,\dots,\dom_K$ where $\dom_m \cap \dom_n = \emptyset$ for $m,n=1,\dots,K$ and $m \neq n$. For a set of spatial kernel functions $\{f_1,\dots,f_L\}$, $\U$ is an orthogonal $p \times p$ joint diagonalization matrix of the matrices $\M_{\dom_k,f_l}(\bs x^{st}(\s))$ for all $k=1,\dots,K$ and $l=1,\dots,L$, which maximizes
\[
\sum_{k=1}^K \sum_{l=1}^L \| \text{diag}(\U \M_{\dom_k,f_l}(\bs x^{st}(\s)) \U^\top )\|_F^2.
\] 
Then, the SNSS.sjd functional is given as $\W (\x) = \U \M_{\dom,f_0}^{-1/2}(\x)$.
\end{defn}

Again, as in the case of the SNSS.jd method, for a finite sample joint diagonalization approximate algorithms need to be used. 

When setting the number of spatial kernel functions $L=1$ and the resulting spatial kernel function to $f=f_0$, then the SNSS.sjd method reduces to the SNSS.jd method. If additionally the spatial domain is only divided into two parts and the transformation step is adapted accordingly, the SNSS.sjd method further reduces to the SNSS.sd method. In similar manner, if the choice of the spatial kernel functions is free but the domain is not partitioned, then the original SBSS method as introduced in \cite{NordhausenOjaFilzmoserReimann2015,BachocGentonNordhausenRuizGazenVirta2020} is obtained. 

Identifiability and affine equivariance results are obtained in the following proposition.

\begin{prop}\label{prop::nssbss.jd2}
The SNSS.sjd functional defined in Definition~\ref{def::nssbss.jd2} is
\begin{description}
\item[(1)] identifiable iff for all pairs $i,j=1,\dots,p$ and $i \neq j$ there exists a pair $k,l$ with $k \in \{1,\dots,K\}$ and $l \in \{1,\dots,L\}$ such that $(\M_{\dom,f_0}^{-1}(\z) \M_{\dom_k,f_l}(\z))_{ii} \neq (\M_{\dom,f_0}^{-1}(\z) \M_{\dom_k,f_l}(\z))_{jj}$,

\item[(2)] affine equivariant as seen in Definition~\ref{def::identifiability}.
\end{description}
\end{prop}

Proposition~\ref{prop::nssbss.jd2} is again more general than Proposition~\ref{prop::nssbss.jd} as more kernel functions can be considered. The most general case is achieved when one member of the set of kernel functions $\{f_1,\dots,f_L\}$ is $f_0$.

\section{Simulations}\label{sec:simulations}

In this part we investigate the performance of the different unmixing matrix estimators which are introduced beforehand in an extensive simulation study. All  simulations are carried out in R version 3.6.1 (\cite{r_software}) with the help of the packages SpatialBSS (\cite{SpatialBSS_package}), JADE (\cite{JADE_package}) and RandomFields (\cite{RandomFields_package}). 

\begin{figure}[t]
\centering
    \begin{minipage}[t]{0.45\textwidth}
        \centering
        \includegraphics[width=\linewidth]{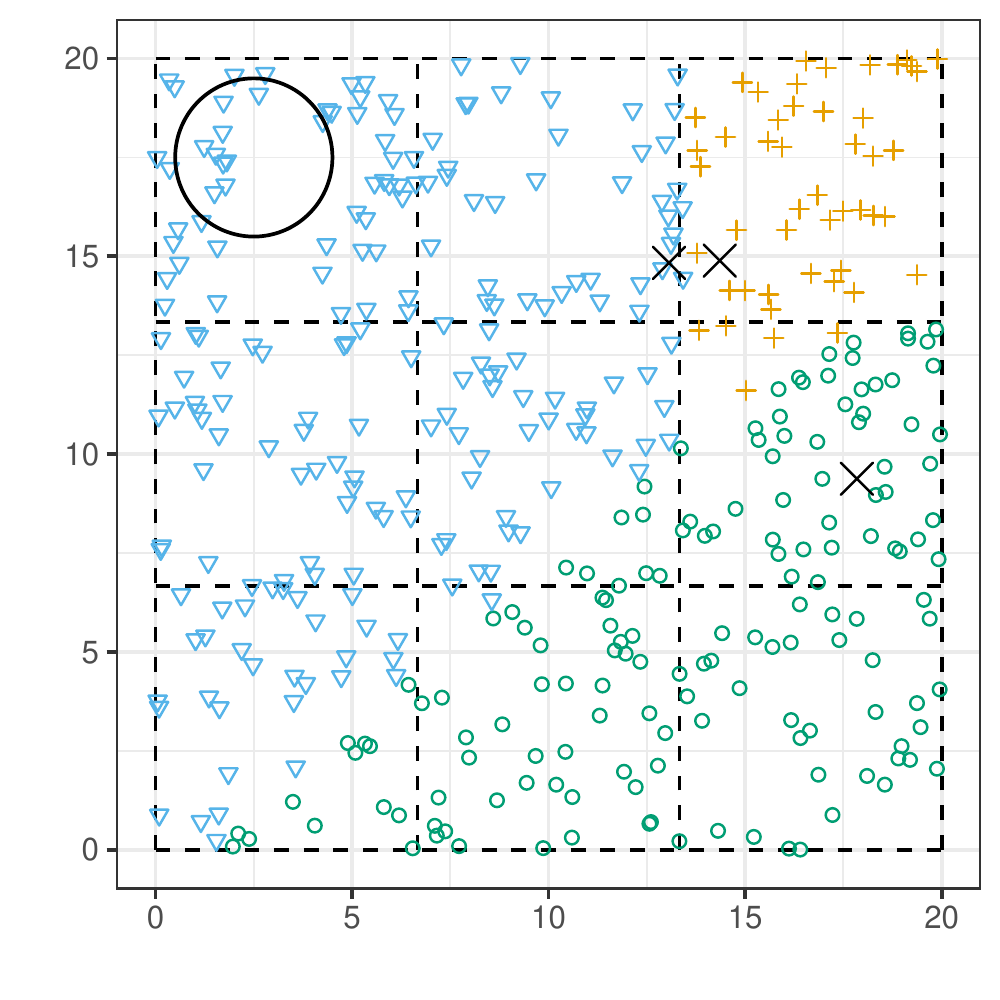}
    \end{minipage}
    \hfill
    \begin{minipage}[t]{0.45\textwidth}
        \centering
        \includegraphics[width=\linewidth]{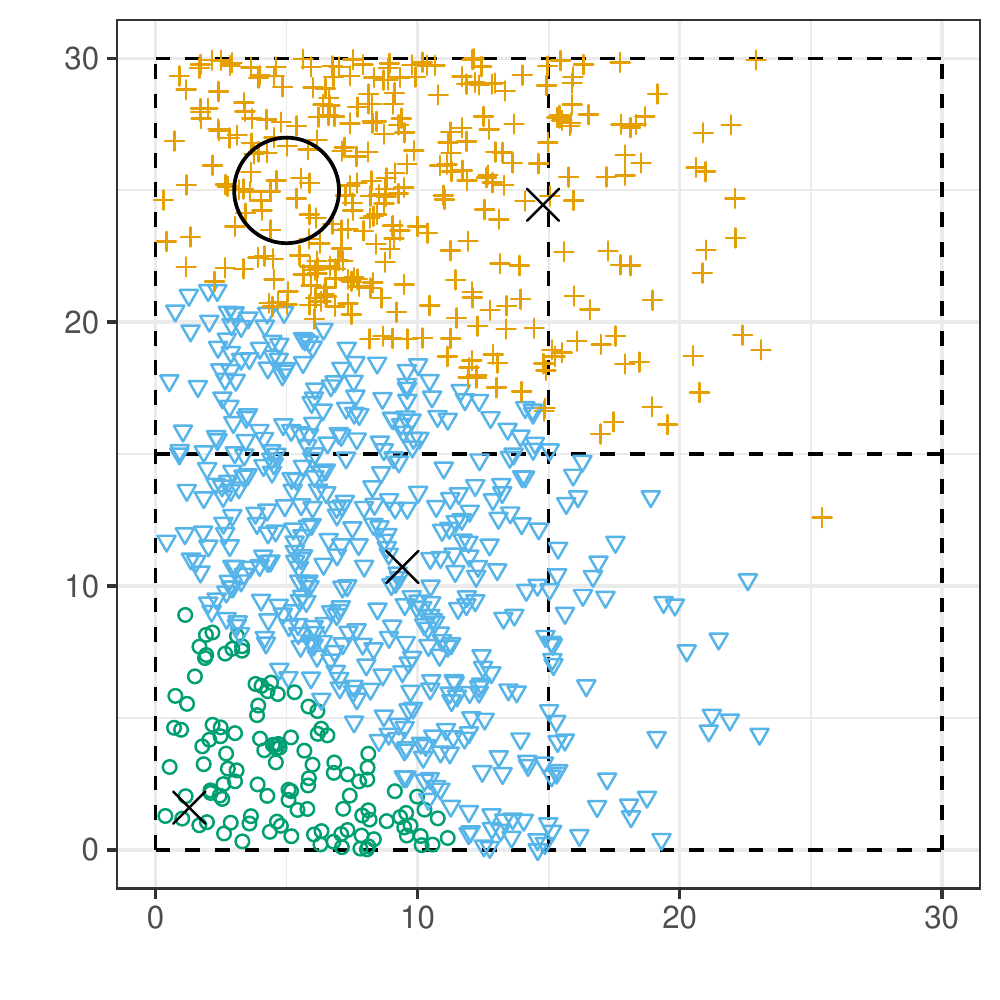}
    \end{minipage}
    \caption{Sample coordinates for a domain size of $20 \times 20$ of the uniform coordinate pattern (left) and for a domain size of $30 \times 30$ for the skewed coordinate pattern (right). The black crosses ($\times$) depict the three randomly placed cluster centers, and the three different colors and shapes hint the corresponding clusters for the sample locations. The dashed lines depict different partitions of the spatial domain. The ring with radius two depicts the parameter used for the spatial kernel functions. }
    \label{fig:sample_locs}
\end{figure}

We use always squared two-dimensional domains of the form $\dom = [0,n] \times [0,n]$ (later denoted also as $n \times n$) where $n \in \{20, 30, 40, 50, 60, 70\}$. The set of sample locations $\loc$ is formed by two different patterns, namely a uniform and skewed pattern. For the uniform coordinate pattern $n^2$ x and y values are sampled from the uniform distribution $U(0,1)$ and then the sampled values are multiplied by $n$. The skewed coordinate pattern is formed by $n^2$ x values that are sampled from the beta distribution $\beta(2,5)$ and $n^2$ y values that are sampled from the uniform distribution $U(0,1)$, again all values are multiplied by $n$. This way of sampling coordinates ensures that the density of sample locations is the same for all domain sizes. In the case of the uniform pattern it equals one throughout the whole domain, whereas the skewed pattern shows more dense sample locations in the left half of the domain. Figure~\ref{fig:sample_locs} depicts one example for the uniform and skewed coordinate pattern for different domain sizes.

Moreover, we randomly divide the spatial domain at hand into three different parts. This is done by randomly placing three locations on the spatial domain that act as cluster centers, which is depicted by the crosses ($\times$) in Figure~\ref{fig:sample_locs}. The three clusters of sample locations are then determined by the lowest Euclidean distance of the sample locations to the cluster centers, this is illustrated by the different colors and shapes for the sample locations in Figure~\ref{fig:sample_locs}.

Using these locations we simulate random fields that follow in all but one setting the SNSS Model (Model~\ref{def::nssbss}). The dimension is set to $p=3$ for all simulations. As our introduced methods are affine equivariant (as seen in Propositions~\ref{prop::nssbss.sd}, \ref{prop::nssbss.jd} and \ref{prop::nssbss.jd2}) we choose without loss of generality $\bs A = \bs I_3$ and $\bs b = \bs 0$ which determines $\x = \z$. The six considered Gaussian distributed random field settings for the latent random field $\z$ are as follows.

\begin{figure}[t]
	\centering
  \includegraphics[width=0.8\linewidth]{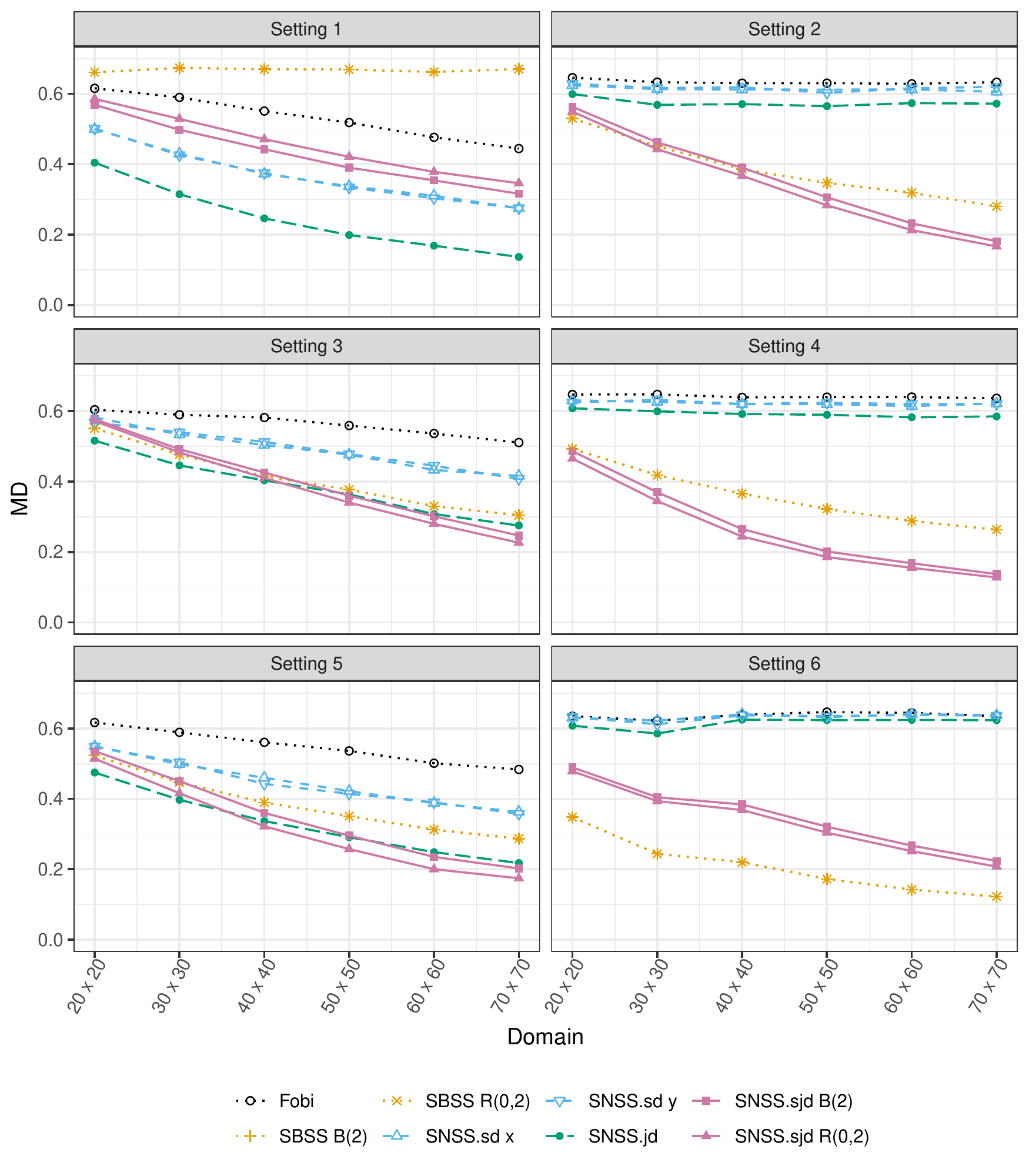}
  \caption{Average MDI based on 2000 simulation repetitions for all random field models, different estimators and sample sizes for the uniform sample location pattern.}
  \label{fig:md_unif}
\end{figure}

\begin{figure}[t]
	\centering
  \includegraphics[width=0.8\linewidth]{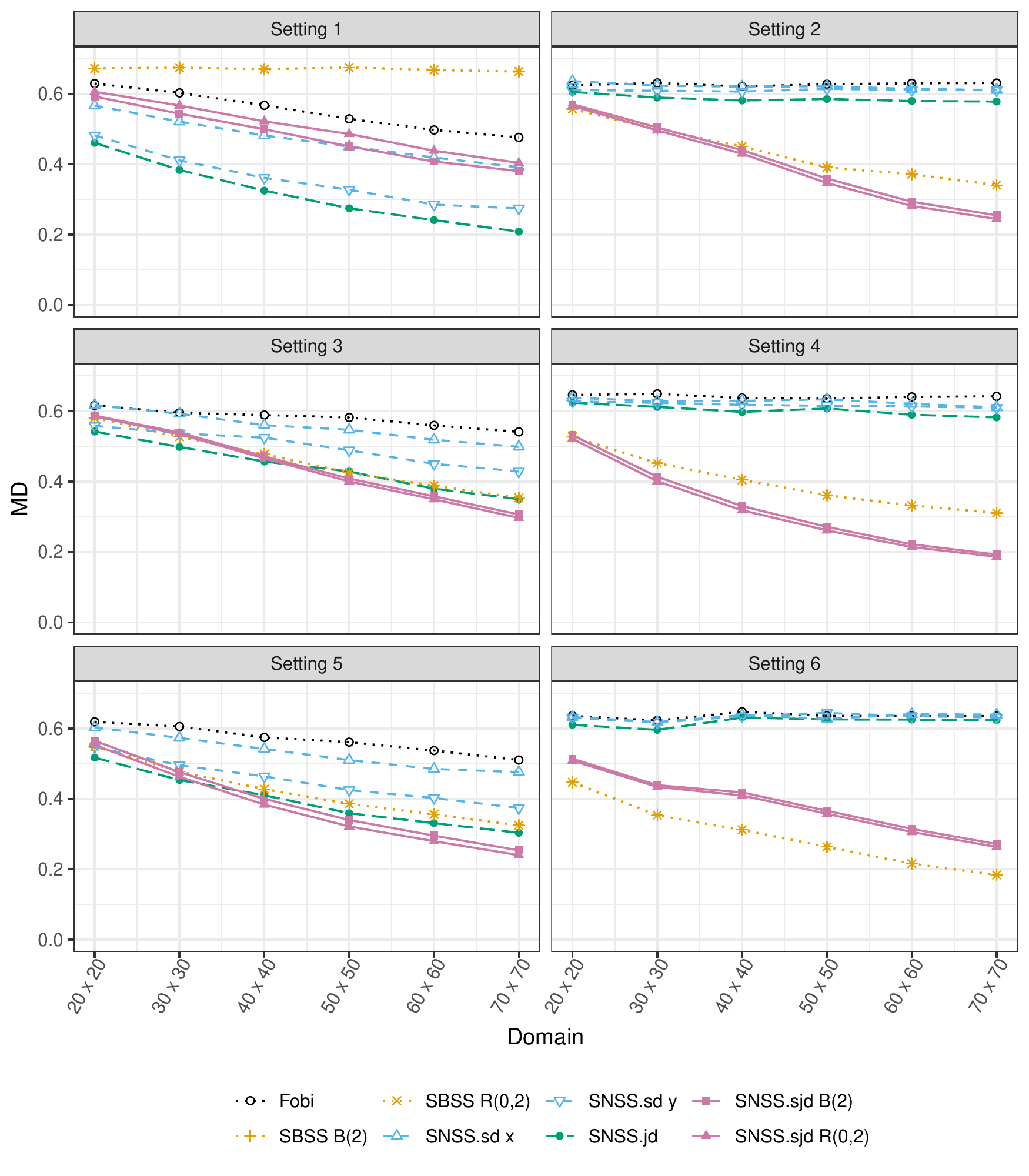}
  \caption{Average MDI based on 2000 simulation repetitions for all random field models, different estimators and sample sizes for the skewed sample location pattern.}
  \label{fig:md_skew}
\end{figure}


\paragraph{Setting 1}
This setting is formed by iid Gaussian distributed $3$-variate random vectors with different covariance matrices in each cluster of sample locations. $\bs \Sigma_{\s}$ equals $\text{diag}(1, 3, 2)$ for cluster one, $\text{diag}(2, 4, 2)$ for cluster two and $\text{diag}(1, 3, 5)$ for cluster three. Thus, $\bs{\Sigma}_{\bs{s} \bs{s'}} = \bs 0$ for the whole spatial domain. 

\paragraph{Setting 2 and 3} 
We sample in each coordinate cluster different random fields independently following the M\'atern covariance function introduced in Section~\ref{sec:intro}. In particular for Setting 2 the covariance function of $z_1(\s)$, $C_{z_1}(h)$ equals $C(h; 1.0, 0.5, 0.5)$ for cluster 1, $C(h; 1.0, 1.0, 1.0)$ for cluster 2 and $C(h; 1.0, 1.0, 2.0)$ for cluster 3. $C_{z_2}(h)$ equals $C(h; 1.0, 1.5, 2.7)$ for cluster 1, $C(h; 1.0, 0.7, 1.0)$ for cluster 2 and $C(h; 1.0, 1.2, 1.9)$ for cluster 3. $C_{z_3}(h)$ equals $C(h; 1.0, 1.2, 1.4))$ for cluster 1, $C(h; 1.0, 0.5, 3.0))$ for cluster 2 and $C(h; 1.0, 0.7, 0.7))$ for cluster 3. Setting 3 is formed in the same fashion as Setting 2 with the only difference that the variance parameters are changed to the ones from Setting 1.

\paragraph{Setting 4 and 5}
These settings are based on the non-stationary extension of the M\'atern covariance function presented in \cite{AnderesStein2011} given by
\[
\begin{split}
C(\s, \s' &;\sigma , \nu, \phi)  = \sigma(\s) \sigma(\s') 
\left( \frac{\phi^2(\s) / 4 \nu(\s)}{\Gamma(\nu(\s)){2 ^ {\nu(\s) - 1}}} \right)^{1/2} \left( \frac{\phi^2(\s') / 4 \nu(\s')}{\Gamma(\nu(\s')){2 ^ {\nu(\s') - 1}}} \right)^{1/2} \\
& \left( \frac{\phi^2(\s)}{8 \nu(\s)} + \frac{\phi^2(\s')}{8 \nu(\s')} \right) ^ {-1}  \left\Vert \tilde{\bs h}\right\Vert ^ {(\nu(\s) + \nu(\s')) / 2}  K_{(\nu(\s) + \nu(\s')) / 2}  \left( \left\Vert \tilde{\bs h} \right\Vert \right) ,  \\ 
& \tilde{\bs h} = \left( \frac{\phi^2(\s)}{8 \nu(\s)} + \frac{\phi^2(\s')}{8 \nu(\s')} \right) ^ {-1/2} (\s - \s'),
\end{split}
\]
where $K_\nu$ is the modified Bessel function of second kind, $\sigma^2:\dom \rightarrow \R ^ +$, $\nu:\dom \rightarrow \R ^ +$ and $\phi:\dom \rightarrow \R ^ +$ are the local variance, shape and range parameter functions. We choose these functions to be of the form $g(\bs x) = \sum_{i=1}^3 c_i \bs 1(\bs x \in \loc_i)$, where $\loc_i$ are the three clusters of sample locations as defined above. The $c_i$ coefficients are the same as the ones from the independently sampled random fields of Setting 2 and 3 for Setting 4 and 5 respectively.

\paragraph{Setting 6}
Setting 6 is a stationary setting, where the entries of the latent field are following a M\'atern covariance function. Explicitly, $C_{z_1}(h)$ equals $C(h; 1.0, 0.5, 1.0)$, $C_{z_2}(h)$ equals $C(h; 1.0, 1.0, 1.5)$ and $C_{z_3}(h)$ equals $C(h; 1.0, 1.5, 2.0)$. 

Note that Setting 1 can be viewed as different white noise for the different clusters of sample locations. For Setting 2 and 3 the random fields are independent between clusters which is not the case for Setting 4 and 5. Setting 2 and 4 have a global constant variance of 1 for all entries of the random field, whereas in Setting 3 and 5 also the variances are different in each cluster of sample locations. Setting 6 is globally stationary with constant variance for each entry of the latent random field. Thus Setting 6 does not really fit into the SNSS framework but is rather into a SBSS framework.

We estimate the unmixing matrix $\hat{\bs W}$ with all SNSS methods described above. For the SNSS.sd method given by Definition~\ref{def::nssbss.sd} we divide the domain in half across the coordinate x axis (SNSS.sd x) and the coordinate y axis (SNSS.sd y). For the SNSS.jd method seen in Definition~\ref{def::nssbss.jd} and SNSS.sjd given by Definition~\ref{def::nssbss.jd2} we define the sub-domains by dividing the domain at hand in four equal squared blocks as shown on the right panel of Figure~\ref{fig:sample_locs}. Additionally, for the SNSS.sjd method we either use a ball kernel with $r=2$ (SNSS.sjd B(2)) and $f_0$ or a ring kernel with $(r_1, r_2) = (0, 2)$ (SNSS.sjd R(0,2)) and $f_0$. This choice keeps the average number of sample locations at $r^2 \pi \approx 12$ for the uniform setting. As contender methods we estimate the unmixing matrix with the SBSS method, introduced in \cite{NordhausenOjaFilzmoserReimann2015, BachocGentonNordhausenRuizGazenVirta2020}, with the same spatial kernel function settings as before but without $f_0$ (SBSS B(2) and SBSS R(0,2)). Lastly, we use the fourth order blind identification (FOBI) method which is a popular independent component analysis (ICA) method that does not utilize spatial information but fourth order cumulants, see \cite{Cardoso1989,NordhausenVirta2019}.

\begin{figure}[t]
	\centering
  \includegraphics[width=0.8\linewidth]{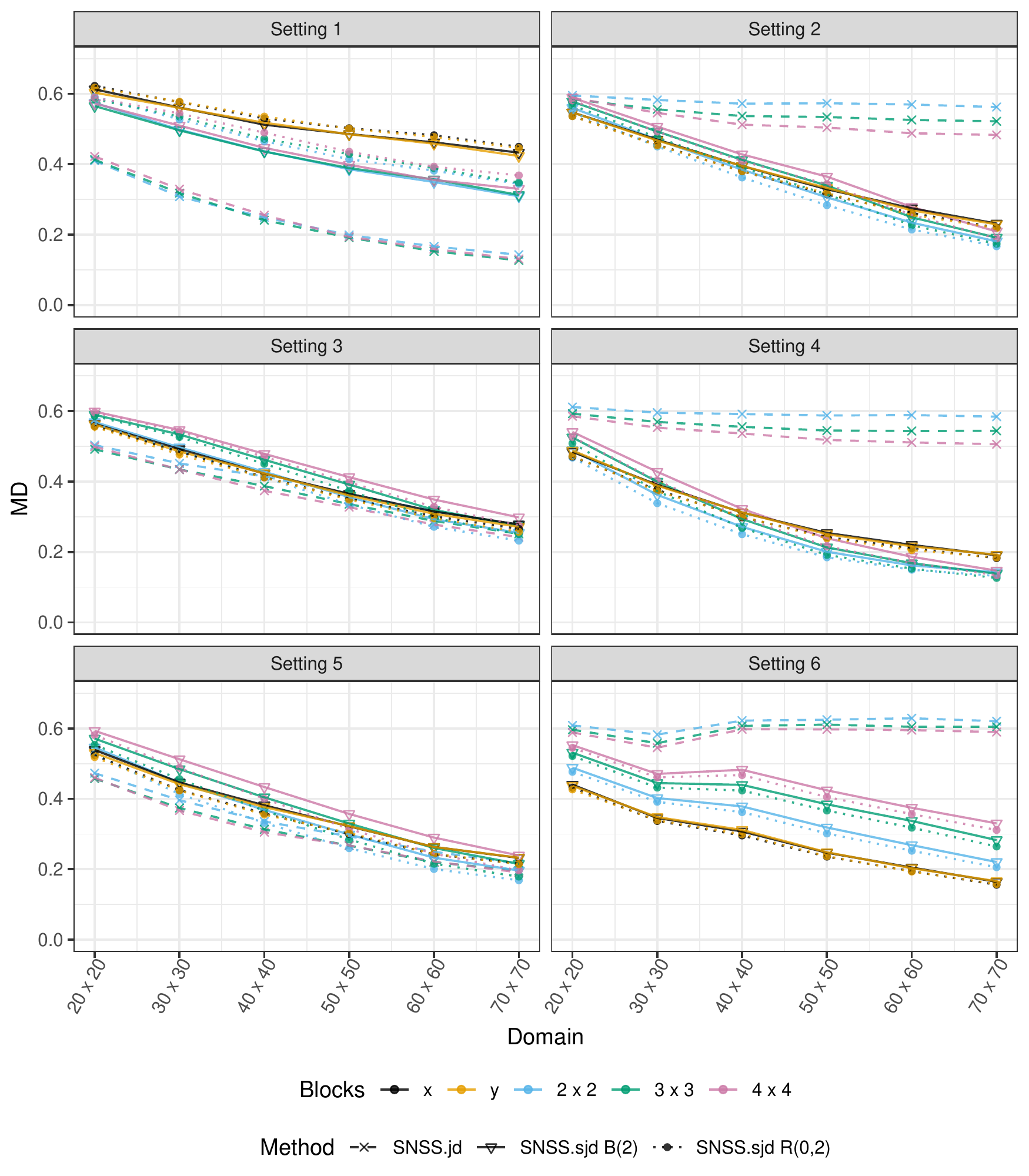}
  \caption{Average MDI based on 2000 simulation repetitions for all random field models, different block sub-domain structures for the SNSS.jd and SNSS.sjd methods and sample sizes for the uniform sample location pattern.}
  \label{fig:MDI_block_unif}
\end{figure}


To evaluate the quality of the unmixing matrix estimate $\hat{\bs W}$ from the different methods we use the minimum distance index (MDI) \cite{IlmonenEtAl2010,LietzenVirtaNordhausenIlmonen2020} which is defined as 
\[
	\text{MDI}(\hat{\bs W} \bs A) = \frac{1}{\sqrt{p-1}} \inf_{\bs J \in \mathcal{J}} \| \bs J \hat{\bs W} \bs A - \bs I_p \|_F .
\]
Here, $\mathcal{J}$ is the set of all matrices that carry exactly one non-zero element in each row and column which corresponds to all matrices of the form $\bs P \bs S \bs D$ that are exactly the indeterminacies of our model definition. The MDI is a function $ \text{MDI}:\R^{p \times p} \rightarrow [0,1]$ where zero indicates that the estimated unmixing matrix meets exactly the real one up to scale, sign and permutation of its rows and one indicates a very poor estimate.

The average MDI based on 2000 simulation iterations for the above estimators for all six considered random field models are presented in Figure~\ref{fig:md_unif} for the uniform sample location pattern. As in Setting 1 the random field shows no spatial dependence all SBSS methods completely fail as they only rely on spatial dependencies and the SNSS.jd method outperforms all contender methods. SNSS.sd is inferior which might be explained by the fact that it only halves the spatial domain, whereas SNSS.jd uses four equally sized sub-domains. Even though the SNSS.sjd methods use the sample covariance matrix inside each sub-domain, additionally (non-informative) local covariance matrices are used which might bring noise into the joint diagonalization algorithm and therefore reduce its performance in this setting. In contrast to Setting 1 only methods that rely on spatial dependencies perform well in Setting 2 and 4 as the variance for this setting equals one for each entry of the random field globally. Interestingly, the SBSS methods still perform well in Setting 2, this might result from the fact that this Setting is based on stationary covariance functions. In Setting 4 SBSS is clearly outperformed by the SNSS.sjd method. As the covariance is non-constant for Setting 3 and 5 also the SNSS.sd and SNSS.jd methods show good performances here. Lastly, as Setting 6 is formed by stationary latent fields with global constant variances, only the SBSS and the SNSS.sjd are expected to deliver meaningful results. However, SBSS shows a better performance because the domain is not split into parts, therefore the effective sample size for the local covariance estimation is higher leading to a better separation. Interestingly, for all simulations where the variance is non-constant FOBI increases its performance as the sample size increases. Also, the choice for the kernel function for SNSS.sjd does not seem to have a high impact.

The results for the skew sample locations pattern are presented in Figure~\ref{fig:md_skew}. The qualitative results are very similar to the uniform setting with two differences. Firstly, the overall performance is worsened for all methods due to the imbalanced distribution of the sample locations. Secondly, the SNSS.sd method where the domain is halved across the y axis clearly increases its performance as the sample locations density is still constant along the y axis. 

The former simulations are carried out for a fixed partition of the spatial domain for the SNSS.jd and SNSS.sjd methods. In this part we investigate the influence of different partitions on the overall performance of the unmixing matrix estimation. We consider sub-divisions into four ($2 \times 2$), nine ($3 \times 3$) and 16 ($4 \times 4$) equally sized squared blocks for both methods. Exemplary, $3 \times 3$ is depicted on the left panel and $2 \times 2$ is depicted on the right panel of Figure~\ref{fig:sample_locs}. Additionally, we half the domain across the x and the y axes for the SNSS.sjd method. The mean MDIs based on 2000 simulation repetitions are shown in Figure~\ref{fig:MDI_block_unif} for the uniform sample location pattern, we do not present the results for the skewed setting as the qualitative results are very similar to the uniform ones. Overall, the influence of the domain sub-division is very minor except for the SNSS.sjd method in Setting 1 and 6. Again, in Setting 6 the performance increases as the sub-division of the domain decreases, and more information is available to estimate the matrices of interest. The optimal case is given when the domain is not divided at all, which leads to the original SBSS method.

Generally, the simulation study showed that SNSS.sjd is a particularly good method as it always improves its performance with increasing sample size, its performance is never among the last and it is among the best in four out of six simulation settings. Therefore, we investigate the usefulness of the SNSS.sjd method on a real data example as follows.

\section{Data example}\label{sec:data_example}

\begin{figure}[t]
\centering
    \begin{minipage}[t]{0.49\textwidth}    
        \centering
        \includegraphics[width=\linewidth]{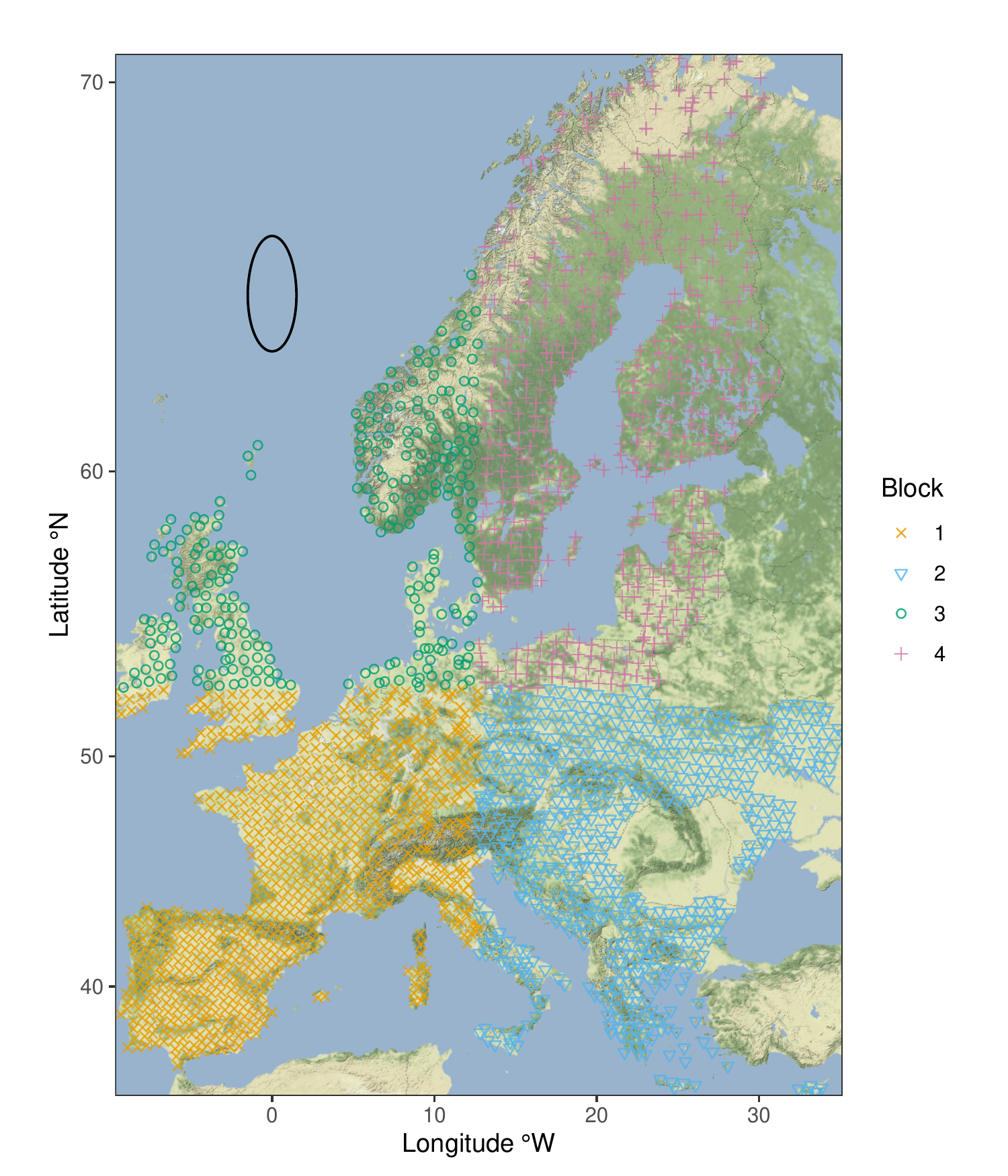}
    \end{minipage}
    \hfill
    \begin{minipage}[t]{0.49\textwidth}
        \centering
        \includegraphics[width=0.83\linewidth]{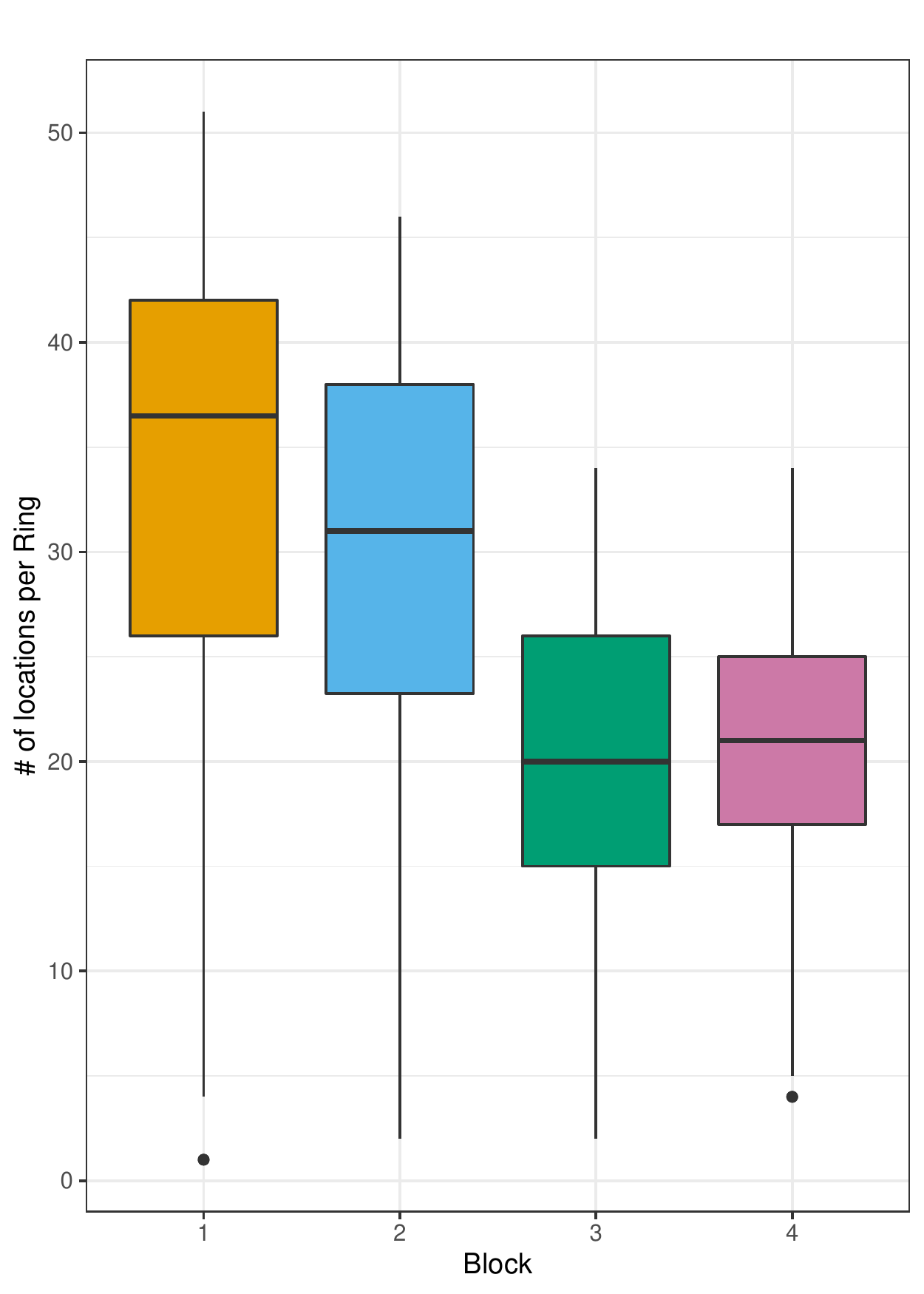}
    \end{minipage}
    \caption{Sample locations for the GEMAS dataset (left panel). The four different colors and shapes illustrate the sub-division of the spatial domain into four equally shaped rectangular blocks. Blocks one to four contain 720, 654, 258 and 475 sample locations respectively. The ring of radius $1.5^{\circ}$ depicts the parameter choice for the ring kernel function. Number of considered neighboring sample locations defined by the ring kernel choice for each of the four blocks (right panel). Map tiles by Stamen Design, under CC BY 3.0. Data by OpenStreetMap, under ODbL.}
    \label{fig:gemas_sample_locs}
\end{figure}

In this section we illustrate the use of the above introduced methods on an environmental application. Specifically, we consider a dataset that is derived from the GEMAS geochemical mapping project \cite{ReimannEtAl2014} which consists of concentration measurements of 18 elements (Al, Ba, Ca, Cr, Fe, K, Mg, Mn, Na, Nb, P, Si, Sr, Ti, V, Y, Zn, Zr) in 2017 agricultural soil samples. This dataset is freely available in the R package robCompositions (\cite{robCompositions_paper}). 

\begin{figure}[t]
\centering
    \begin{minipage}[t]{0.49\textwidth}    
        \centering
        \includegraphics[width=\linewidth]{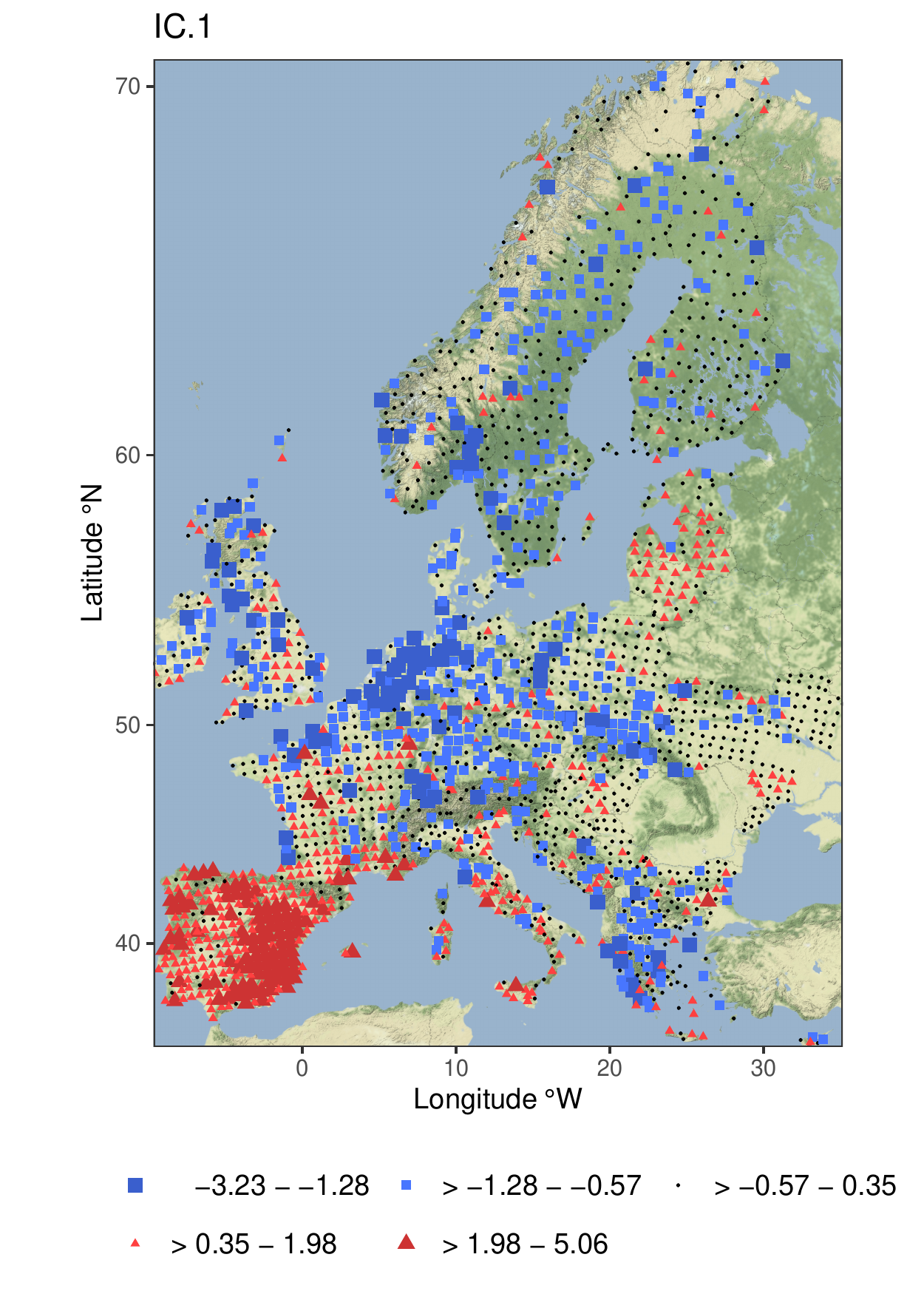}
    \end{minipage}
    \hfill
    \begin{minipage}[t]{0.49\textwidth}
        \centering
        \includegraphics[width=\linewidth]{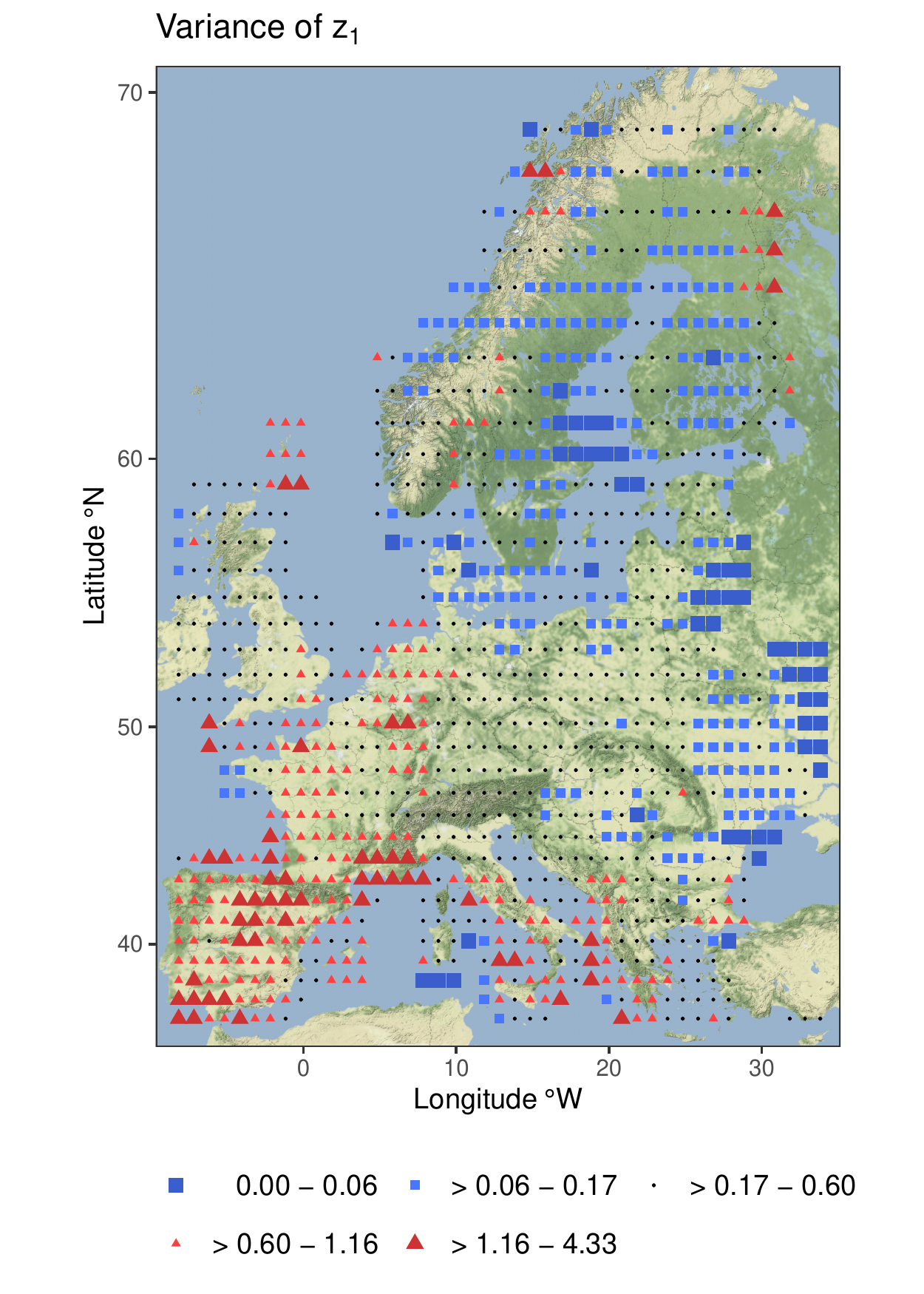}
    \end{minipage}
    \caption{Map of the first entry of the estimated latent field (left) and its corresponding moving block variance map (right). Map tiles by Stamen Design, under CC BY 3.0. Data by OpenStreetMap, under ODbL.}
    \label{fig:IC1}
\end{figure}

\begin{figure}[t]
\centering
    \begin{minipage}[t]{0.49\textwidth}    
        \centering
        \includegraphics[width=\linewidth]{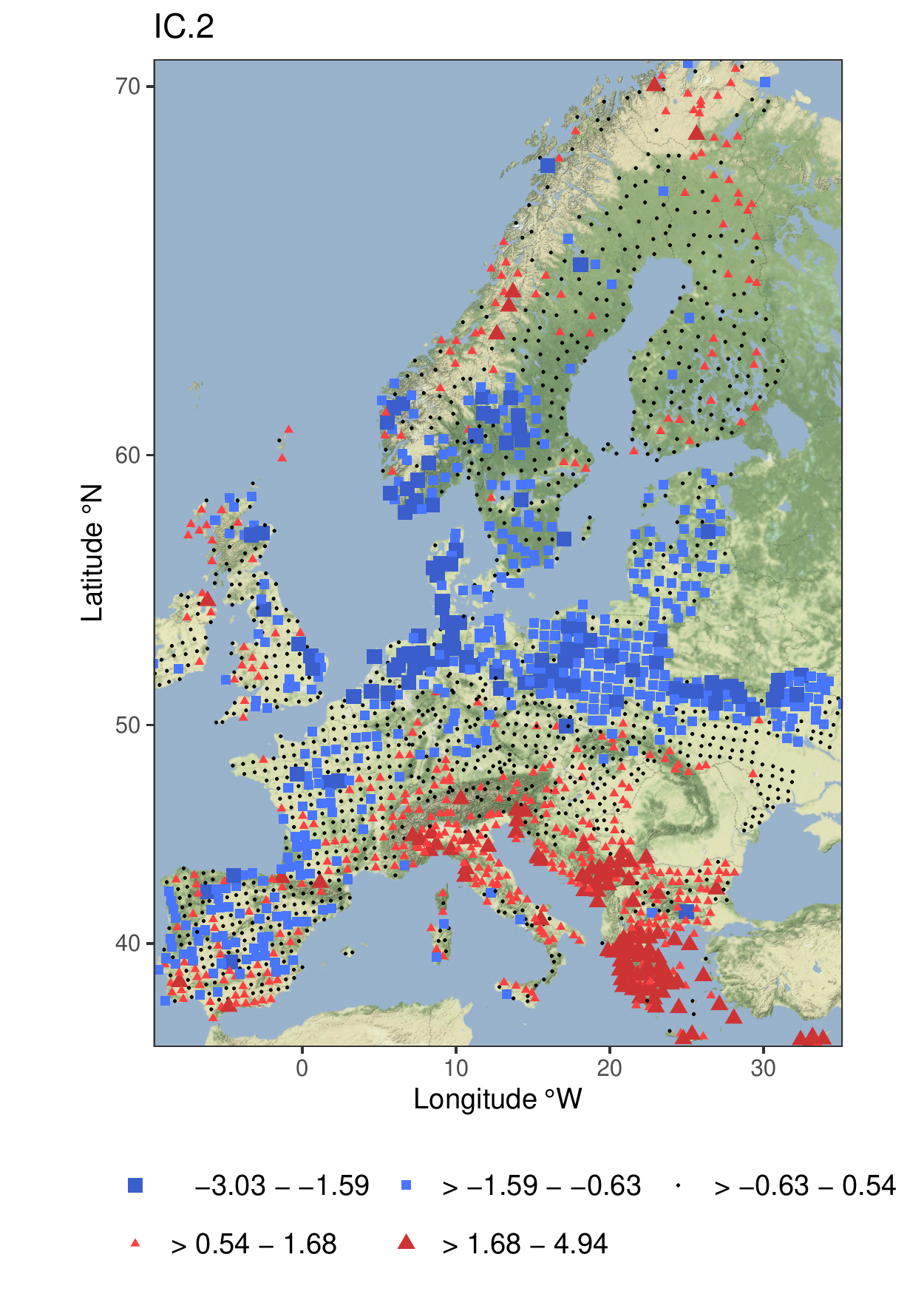}
    \end{minipage}
    \hfill
    \begin{minipage}[t]{0.49\textwidth}
        \centering
        \includegraphics[width=\linewidth]{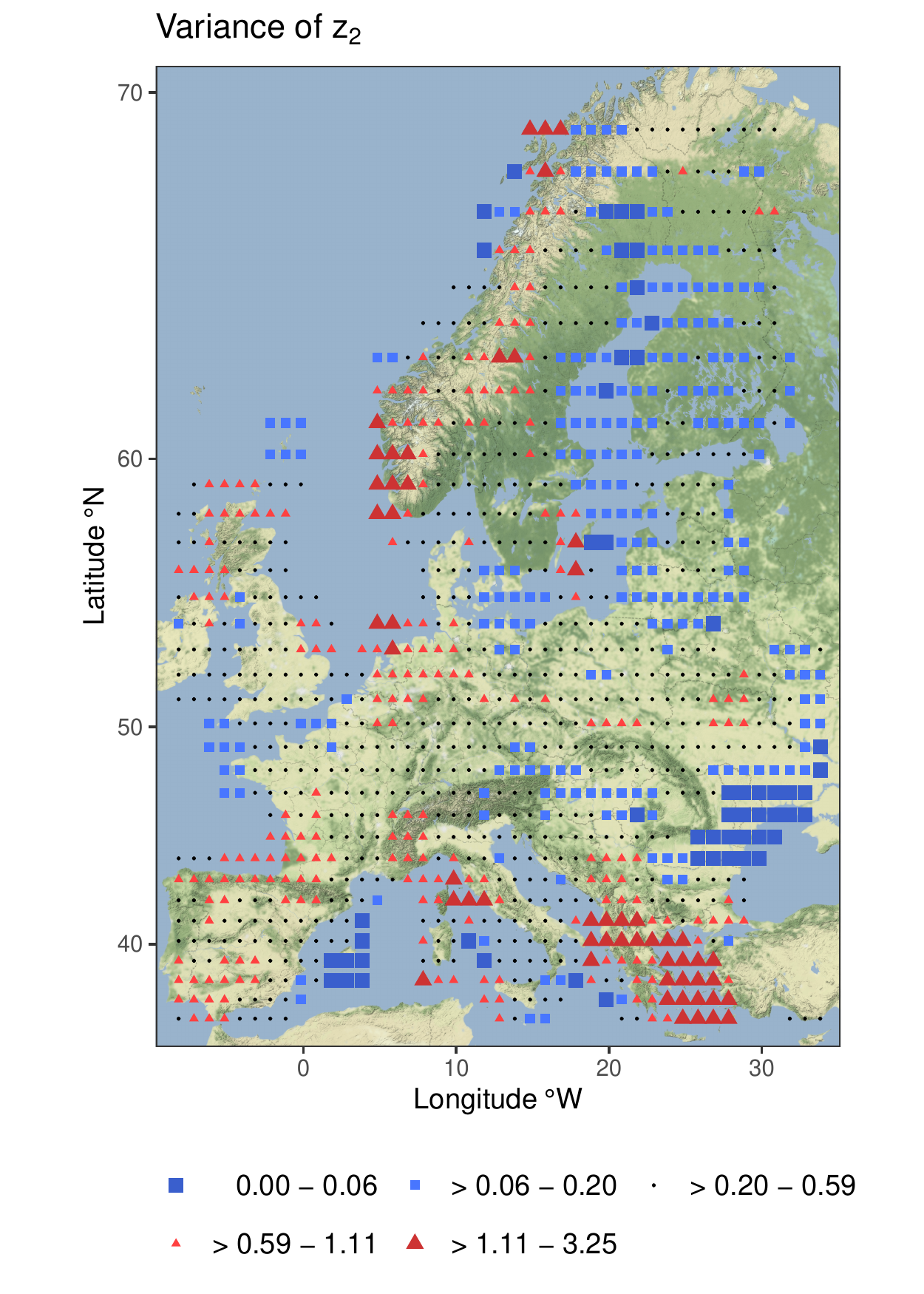}
    \end{minipage}
    \caption{Map of the second entry of the estimated latent field (left) and its corresponding moving block variance map (right). Map tiles by Stamen Design, under CC BY 3.0. Data by OpenStreetMap, under ODbL.}
    \label{fig:IC2}
\end{figure}

As it is common practice in geochemical applications we respect the relative information of the data by performing typical compositional data analysis transformations prior the actual SNSS analysis. In a BSS context this is for example discussed in \cite{muehlmann2020independent,NordhausenOjaFilzmoserReimann2015} and we follow in the exact same fashion as outlined in \cite{NordhausenOjaFilzmoserReimann2015}. We first perform an isometric log-ratio (ilr) transformation by using pivot coordinates, and then apply the SNSS method. The loadings matrix is formed by combining the contrast matrix and the estimated unmixing matrix. Here the contrast matrix is an orthogonal matrix that transforms the data from centered log-ratio (clr) into ilr coordinates. Details on clr, ilr and compositional data analysis in general are given for example in \cite{comp_book}. Note that the ilr transformation reduces the dimension of the dataset by one, therefore $p=17$. 

We carry out SNSS.sjd as it has the overall best performance in the simulation study above. We divide the domain into four equally sized rectangles where the four resulting blocks of sample locations are depicted in the left panel of Figure~\ref{fig:gemas_sample_locs}. The circle on that Figure illustrates the parameter $(r_1, r_2) = (0^{\circ}, 1.5^{\circ})$ for the used ring kernel function and the right panel of Figure~\ref{fig:gemas_sample_locs} shows boxplots of the number of neighboring sample locations defined by the ring kernel choice for each of the four considered blocks of sample locations. Additional to the ring kernel function choice we also include the covariance matrix for each of the four blocks (kernel function $f_0$). 

We compute moving block variance maps for each entry of the latent random field, to hint the possible non-stationary variances. Specifically, we overlaid the domain by a grid with a resolution of one degree where the center is placed on the minimum longitude and latitude value present in the dataset. The variance for each cell of the grid is computed by considering all sample locations that lie inside a block of size $3^{\circ} \times 3^{\circ}$ that is placed on that cell. 

\begin{table}[t]
\centering
\caption{Values of the combined loadings matrix that transforms the clr data into the first two components of the estimated latent random field.}
\begin{tabular}{lcclcc}
  \toprule
 & $z_1$ & $z_2$ & &  $z_1$ & $z_2$ \\ 
  \midrule
clr(Al) & \textbf{1.72} & \textbf{1.47} &  clr(Nb) & -0.50 & 0.02 \\ 
clr(Ba) & -0.23 & -0.29 &clr(P) & -0.15 & -0.40 \\  
clr(Ca) & 0.06 & 0.09 &  clr(Si) & 0.71 & -0.32 \\ 
clr(Cr) & -0.55 & \textbf{1.01} & clr(Sr) & 0.42 & -0.15 \\ 
clr(Fe) & 0.86 & -0.18 & clr(Ti) & \textbf{1.38} & -0.12 \\ 
clr(K) & 0.13 & -0.27 &  clr(V) & \textbf{-1.29} & -0.68 \\ 
clr(Mg) & 0.30 & -0.17 & clr(Y) & -0.05 & 0.53 \\ 
clr(Mn) & -0.36 & 0.26 & clr(Zn) & -0.60 & -0.06 \\
clr(Na) & \textbf{-1.22} & 0.18 & clr(Zr) & -0.63 & \textbf{-0.91} \\
   \bottomrule
\end{tabular}
\label{tab:loadings}
\end{table}

After visual inspection of all recovered entries of the latent random field and the corresponding moving block variance maps we exemplary present the first two entries in Figure~\ref{fig:IC1} and \ref{fig:IC2}. The corresponding combined loadings (matrix product of the contrast and the estimated unmixing matrix) that transform the clr data into the first and second entry of the latent random field are presented in Table~\ref{tab:loadings}. A cluster of high values for the first component of the latent random field is found on the Iberian Peninsula. This cluster is mostly formed by the high balance between the pair Al, Ti and Na, V as the corresponding loadings show roughly equal values with opposite signs. The second component of the latent random field shows a cluster of high variance as well as high values in Greece, along the Balkan up to the northern and central part of Italy. The high loading of clr(Al) and the roughly equal absolute values of the clr(Cr) and clr(Zr) loadings suggest that this entry is mostly driven by a positive log-ratio between Cr and Zr combined with the high relative dominance of Al. The opposite effect is observed for the cluster of low values from mid to east Europe and the southern part of Scandinavia. Deeper investigation of the found latent random field and the possible driving physical phenomena can be achieved by geological experts.

\section{Conclusion}\label{sec:conclusion}

BSS has been successfully used in many scientific applications \cite{comon2010handbook}. BSS has a long tradition for iid data where it is known as independent component analysis (ICA) and for stationary and non-stationary time series \cite{PanMatilainenTaskinenNordhausen2021}.
Recently BSS approaches were suggested  for stationary spatial data  \cite{NordhausenOjaFilzmoserReimann2015,BachocGentonNordhausenRuizGazenVirta2020}. In this paper, we combine ideas from non-stationary time series methods and spatial stationary BSS to develop approaches for non-stationary spatial data. We formulate a spatial non-stationary blind source separation model and provide three different estimators that are based on the joint diagonalization of covariance and local covariance matrices for sub-divisions of the spatial domain. These estimators can be easily applied on spatial datasets with irregular sample locations and their use is illustrated in an extensive simulation study and on an environmental application. 

Interesting future research would be to derive asymptotic results for the different estimators. Furthermore, it is of great interest to explore the use of the SNSS methods in the context of spatial prediction. The entries of the latent random field are uncorrelated, therefore, $p$ univariate non-stationary models can be built which is much simpler as building one multivariate model for the original data. In the stationary case, such an approach seemed promising as discussed in \cite{muehlmann2020cokriging}. 
Another interesting question would be to test if all latent components are actually informative and non-stationary, perhaps some exhibit spatial dependence but are stationary and others might be just white noise. In such cases modelling could be simplified. The separation of stationary from white noise processes in SBSS is for example discussed in \cite{muehlmann2020test}.
We have focused so far on simple rectangular subdivisions of the domain at hand for the SNSS estimators, but irregular divisions might also be beneficial.

In a time series context \cite{PfisterWeichwaldBuhlmannScholkopf2019} viewed such a partition of the data as a realization of grouped data and adapted the BSS model to such a case. A motivating example would be EEG signals where the sensors are placed on the same locations for different patients ensuring the same way of mixing. The measurements for each patient then form the different groups. However, motivation for the adaptation to the spatial setting is a future problem.


\section*{Acknowledgement}

The work of CM and KN was supported by the Austrian Science Fund P31881-N32.

\section*{Appendix}

\begin{pf_1} 

Identifiability: For a given simultaneous diagonalizer $\W$ the first optimization equation writes
\[
\begin{split} 
\bs I_p & = \W \M_{\dom_1,f_0}(\x) \W^\top = \W \A \A^{-1} \M_{\dom_1,f_0}(\x) \A^{- \top} \A^\top \W^\top \\
& = \W \A \M_{\dom_1,f_0}(\z) \A^\top \W^\top.
\end{split} 
\] 

As $\M_{\dom_1,f_0}(\z)$ is a diagonal matrix with strictly positive diagonal elements by assumption it follows that $\W \A \M_{\dom_1,f_0}^{1/2}(\z) = \V$ where $\V$ is a $p \times p$ orthogonal matrix. With that the second optimization equations writes as
\[ 
\bs D_{\dom_1 \dom_2} = \W \A \M_{\dom_2,f_0}(\z) \A^\top \W^\top = \V \M_{\dom_1,f_0}^{-1}(\z) \M_{\dom_2,f_0}(\z) \V^\top.
\] 

$\Leftarrow$: As the diagonal elements of the matrix $\M_{\dom_1,f_0}^{-1}(\z) \M_{\dom_2,f_0}(\z)$ are pairwise distinct the matrix $\bs D_{\dom_1 \dom_2}$ has $p$ unique one-dimensional eigenspaces that are orthogonal. Therefore, $\V$ can only be of the form $\bs P \bs S$, and hence $\W \A  = \bs P \bs S \M_{\dom_1,f_0}(\z)^{-1/2}$  which is of the form $\bs P \bs S \bs D$.

$\Rightarrow$: Assume w.l.o.g. that the first two diagonal elements of $\M_{\dom_1,f_0}^{-1}(\z) \M_{\dom_2,f_0}(\z)$ are equal, denoted as $\lambda$. Then from the second optimization equation the first two eigenvalue equations write $\bs D_{\dom_1 \dom_2} \bs v_1 = \bs v_1 \lambda$ and $\bs D_{\dom_1 \dom_2} \bs v_2 = \bs v_2 \lambda$ where the eigenvectors can be written as $\bs v_1 =(1/\sqrt 2, 1/\sqrt 2,0,\dots,0)^ \top$ and $\bs v_2 = (1/\sqrt 2, - 1/\sqrt 2,0,\dots,0)^ \top$. But then $\V$ is not of the form $\bs P \bs S$ and consequently $\W \A$ is not of the form $\bs P \bs S \bs D$.

Affine equivariance: Consider an affine transformation of $\x$ written as $\bs x^* (\s) = \bs B \x + \bs a$, where $\bs B$ is an invertible $p \times p$ matrix. The unmixing matrix functional $\W^* = \W^*(\bs x^* (\s))$ satisfies  
\[
\W^* \M_{\dom_1,f_0}(\bs x^* (\s)) \W^{*\top} = \bs I_p, \W^* \M_{\dom_2,f_0}(\bs x^* (\s)) \W^{*\top} = \bs D^*_{\dom_1 \dom_2},
\] 

for a diagonal matrix $\bs D^*_{\dom_1 \dom_2}$. But because of the affine equivariance of local covariance matrices it also follows that
\[
\W^* \bs B \M_{\dom_1,f_0}(\x) \bs B^\top \W^{*\top} = \bs I_p, \W^* \bs B \M_{\dom_2,f_0}(\x) \bs B^\top \W^{*\top} = \bs D^*_{\dom_1 \dom_2}.
\] 

From the last equations $\W^* \bs B$ can be identified as the unmixing matrix $\W(\x)$, this leads to $\W^*(\bs x^* (\s)) = \W(\x) \bs B^{-1}$ which concludes the proof. \qed 
\end{pf_1}

\begin{pf_2} Definition~\ref{def::nssbss.jd} is a special case of Definition~\ref{def::nssbss.jd2}, therefore, the proof of Proposition~\ref{prop::nssbss.jd} is a special case of the proof of Proposition~\ref{prop::nssbss.jd2} for $L=1$. \qed

\end{pf_2}

\begin{pf_3}
Identifiability: For a given unmixing matrix $\W = \U \M_{\dom,f_0}^{-1/2}(\x)$ from the transformation step it follows that
\[ 
\begin{split}
\bs I_p & = \U \bs I_p \U^\top = \U \M_{\dom,f_0}^{-1/2}(\x) \M_{\dom,f_0}(\x ) \M_{\dom,f_0}^{-1/2}(\x) \U^\top \\
& = \W \A \M_{\dom,f_0}(\z ) \A^\top \W^\top.
\end{split}
\]

As $\M_{\dom,f_0}(\z)$ is a diagonal matrix with strictly positive diagonal elements by assumption it follows that $\W \A \M_{\dom,f_0}^{1/2}(\z) = \V$ where $\V$ is a $p \times p$ orthogonal matrix. From the maximization equation it follows that
\[
\begin{split}
& \sum_{k=1}^K \sum_{l=1}^L \| \text{diag}(\U \M_{\dom_k,f_l}(\bs x^{st}(\s))) \U^\top \|_F^2 \\
= & \sum_{k=1}^K \sum_{l=1}^L (\| \U \M_{\dom_k,f_l}(\bs x^{st}(\s)) \U^\top \|_F^2 - \| \text{off}(\U \M_{\dom_k,f_l}(\bs x^{st}(\s))) \U^\top \|_F^2 ),
\end{split}
\] 
here $\text{off}(\cdot)$ is obtained by setting all off-diagonal elements of the squared-matrix argument to zero. We have $\M_{\dom,f_0}^{-1/2}(\x) = \bs O \M_{\dom,f_0}^{-1/2}(\z) \A^{-1}$, with a unique orthogonal matrix $\bs O$, from \cite[Theorem 2.1]{IlmonenOjaSerfling2012}. Hence, one can show that there is an orthogonal matrix $\U'$ such that $\U' \M_{\dom_k,f_l}(\bs x^{st}(\s))) \U'^\top$, $k=1,\dots,K$, $l=1,\dots,L$ are diagonal (see the equivariance proof below). As $\U$ maximizes the sum of Frobenius norms of the diagonals we have that $\U \M_{\dom_k,f_l}(\bs x^{st}(\s))) \U^\top$, $k=1,\dots,K$, $l=1,\dots,L$ are diagonal. But also
\[
\begin{split}
\U \M_{\dom_k,f_l}(\bs x^{st}(\s)) \U^\top = 
 \W \A \M_{\dom_k,f_l}(\z) \A^\top \W^\top 
=  \V \M_{\dom,0}^{-1}(\z) \M_{\dom_k,f_l}(\z) \V^\top.
\end{split}
\] 

Therefore, all matrices $\V \M_{\dom,f_0}^{-1}(\z) \M_{\dom_k,f_l}(\z) \V^\top$ for $k=1,\dots,K$ and $l=1,\dots,L$ are diagonal.

$\Leftarrow$: For all pairs $i,j=1,\dots,p$ and $i \neq j$ there exists a pair $k,l$ with $k \in \{1,\dots,K\}$ and $l \in \{1,\dots,L\}$ such that $(\M_{\dom,f_0}^{-1}(\z) \M_{\dom_k,f_l}(\z))_{ii} \neq (\M_{\dom,f_0}^{-1}(\z) \M_{\dom_k,f_l}(\z))_{jj}$. Hence, only choices of $\V = \bs P \bs S$ keep all matrices $\V \M_{\dom,f_0}^{-1}(\z) \M_{\dom_k,f_l}(\z) \V^\top$ for $k=1,\dots,K$ and $l=1,\dots,L$ diagonal. This is for instance shown in \cite{BachocGentonNordhausenRuizGazenVirta2020}. Therefore, $\W \A  = \bs P \bs S \M_{\dom,f_0}^{-1/2}(\z)$  which is of the form $\bs P \bs S \bs D$.

$\Rightarrow$: Assume that there exists one pair $i,j \in \{1,\dots,p\}$ with $i \neq j$ where for all pairs $k,l$ with $k=1,\dots,K$ and $l=1,\dots,L$, it holds that $(\M_{\dom,f_0}^{-1}(\z) \M_{\dom_k,f_l}(\z))_{ii} = (\M_{\dom,f_0}^{-1}(\z) \M_{\dom_k,f_l}(\z))_{jj}$. W.l.o.g assume that $i=1$ and $j=2$ then $\V$ could be a block diagonal matrix with the first block $((1/\sqrt 2, 1/\sqrt 2,)^\top, (1/\sqrt 2, -1/\sqrt 2,)^\top)$ and the second block $\bs I_{p-2}$. This choice of $\V$ still keeps all matrices $\V \M_{\dom,f_0}^{-1}(\z) \M_{\dom_k,f_l}(\z) \V^\top$ for $k=1,\dots,K$ and $l=1,\dots,L$ diagonal. But then $\V$ is not of the form $\bs P \bs S$ and consequently $\W \A$ is not of the form $\bs P \bs S \bs D$.

Affine equivariance: Consider an affine transformation of $\x$ written as $\bs x^* (\s) = \bs B \x + \bs c$, where $\bs B$ is an invertible $p \times p$ matrix. From \cite{IlmonenOjaSerfling2012} Theorem 2.1 it follows that $\M_{\dom,f_0}^{-1/2}(\bs x^* (\s)) = \V  \M_{\dom,f_0}^{-1/2}(\x) \bs B^{-1}$, where $\V$ is a unique $p \times p$ orthogonal matrix. The unmixing matrix functional $\W^*(\bs x^* (\s)) = \U^* \M_{\dom,f_0}^{-1/2}(\bs x^* (\s))$ maximizes
\[
\begin{split}
& \sum_{k=1}^K \sum_{l=1}^L \| \text{diag}(\U^* \M_{\dom_k,f_l}(\bs x^{st*}(\s))) \U^{*\top}) \|_F^2 \\ 
 = & \sum_{k=1}^K \sum_{l=1}^L \| \text{diag}(\U^* \M_{\dom,f_0}^{-1/2}(\bs x^* (\s)) \M_{\dom_k,f_l}(\bs x^{*}(\s))) \M_{\dom,f_0}^{-1/2}(\bs x^* (\s)) \U^{*\top}) \|_F^2 \\
 = & \sum_{k=1}^K \sum_{l=1}^L \| \text{diag}(\U^* \V  \M_{\dom,f_0}^{-1/2}(\x) \bs B^{-1} \bs B \M_{\dom_k,f_l}(\x) \bs B^{\top} \bs B^{-\top} \M_{\dom,f_0}^{-1/2}(\x) \V^\top \U^{*\top}) \|_F^2 \\
 = & \sum_{k=1}^K \sum_{l=1}^L \| \text{diag}(\U^* \V   \M_{\dom_k,f_l}(\bs x^{st}(\s))  \V^\top \U^{*\top}) \|_F^2.
\end{split}
\] 

Therefore, $\U = \U^* \V$ is the joint diagonalizer of the matrices $\M_{\dom_k,f_l}(\bs x^{st}(\s))$, $k=1,\dots,K$, $l=1,\dots,L$. This leads to
\[
\begin{split}
\W^*(\bs x^* (\s)) & = \U^* \M_{\dom,f_0}^{-1/2}(\bs x^* (\s)) =
\U \V^\top  \V  \M_{\dom,f_0}^{-1/2}(\x) \bs B^{-1} = \U \M_{\dom,f_0}^{-1/2}(\x) \bs B^{-1} \\
& = \W(\x) \bs B^{-1},
\end{split}
\]

which concludes the proof. \qed

\end{pf_3}

\bibliographystyle{agsm}


\begin{thebibliography}{10}
\expandafter\ifx\csname url\endcsname\relax
  \def\url#1{\texttt{#1}}\fi
\expandafter\ifx\csname urlprefix\endcsname\relax\def\urlprefix{URL }\fi
\expandafter\ifx\csname href\endcsname\relax
  \def\href#1#2{#2} \def\path#1{#1}\fi

\bibitem{GuttorpGneiting2006}
P.~Guttorp, T.~Gneiting, {Studies in the History of Probability and Statistics
  XLIX on the Matérn Correlation Family}, Biometrika 93~(4) (2006) 989--995.
\newblock \href {https://doi.org/https://doi.org/10.1093/biomet/93.4.989}
  {\path{doi:https://doi.org/10.1093/biomet/93.4.989}}.

\bibitem{GentonKleiber2015}
M.~G. Genton, W.~Kleiber, {Cross-Covariance Functions for Multivariate
  Geostatistics}, Statistical Science 30~(2) (2015) 147 -- 163.
\newblock \href {https://doi.org/10.1214/14-STS487}
  {\path{doi:10.1214/14-STS487}}.

\bibitem{GoulardVoltz1992}
M.~Goulard, M.~Voltz, Linear coregionalization model: Tools for estimation and
  choice of cross-variogram matrix, Mathematical Geology 24 (1992) 269--286.
\newblock \href {https://doi.org/10.1007/BF00893750}
  {\path{doi:10.1007/BF00893750}}.

\bibitem{Wackernagel2003}
H.~Wackernagel, Multivariate Geostatistics, Springer, 2003.

\bibitem{GneitingKleiberSchlather2010}
T.~Gneiting, W.~Kleiber, M.~Schlather, Matern cross-covariance functions for
  multivariate random fields, Journal of the American Statistical Association
  105 (2010) 1167--1177.
\newblock \href {https://doi.org/10.1198/jasa.2010.tm09420}
  {\path{doi:10.1198/jasa.2010.tm09420}}.

\bibitem{Sampson2010}
P.~D. Sampson, Constructions for Nonstationary Spatial Processes, CRC Press,
  2010, pp. 119--130.
\newblock \href {https://doi.org/10.1201/9781420072884-c9}
  {\path{doi:10.1201/9781420072884-c9}}.

\bibitem{AnderesStein2011}
E.~B. Anderes, M.~L. Stein, Local likelihood estimation for nonstationary
  random fields, Journal of Multivariate Analysis 102~(3) (2011) 506 -- 520.
\newblock \href {https://doi.org/https://doi.org/10.1016/j.jmva.2010.10.010}
  {\path{doi:https://doi.org/10.1016/j.jmva.2010.10.010}}.

\bibitem{VuZammitMangionCressie2021}
Q.~Vu, A.~Zammit-Mangion, N.~Cressie, Modeling nonstationary and asymmetric
  multivariate spatial covariances via deformations, arXiv (2021) 2004.08724.

\bibitem{GelfandSchmidtBanerjeeSirmans2004}
A.~E. Gelfand, A.~M. Schmidt, S.~Banerjee, C.~F. Sirmans, Nonstationary
  multivariate process modeling through spatially varying coregionalization,
  Test 13 (2004) 263--312.
\newblock \href {https://doi.org/10.1007/BF02595775}
  {\path{doi:10.1007/BF02595775}}.

\bibitem{KleiberNychka2012}
W.~Kleiber, D.~Nychka, Nonstationary modeling for multivariate spatial
  processes, Journal of Multivariate Analysis 112 (2012) 76--91.
\newblock \href {https://doi.org/10.1016/j.jmva.2012.05.011}
  {\path{doi:10.1016/j.jmva.2012.05.011}}.

\bibitem{NordhausenOjaFilzmoserReimann2015}
K.~Nordhausen, H.~Oja, P.~Filzmoser, C.~Reimann, Blind source separation for
  spatial compositional data, Mathematical Geosciences 47~(7) (2015) 753--770.
\newblock \href {https://doi.org/https://doi.org/10.1007/s11004-014-9559-5}
  {\path{doi:https://doi.org/10.1007/s11004-014-9559-5}}.

\bibitem{BachocGentonNordhausenRuizGazenVirta2020}
F.~Bachoc, M.~G. Genton, K.~Nordhausen, A.~Ruiz-Gazen, J.~Virta, {Spatial blind
  source separation}, Biometrika 107~(3) (2020) 627--646.
\newblock \href {https://doi.org/10.1093/biomet/asz079}
  {\path{doi:10.1093/biomet/asz079}}.

\bibitem{comon2010handbook}
P.~Comon, C.~Jutten, Handbook of Blind Source Separation: Independent Component
  Analysis and Applications, Academic Press, Amsterdam, 2010.

\bibitem{NordhausenOja2018}
K.~Nordhausen, H.~Oja, Independent component analysis: A statistical
  perspective, WIREs: Computational Statistics 10 (2018) e1440.
\newblock \href {https://doi.org/10.1002/wics.1440}
  {\path{doi:10.1002/wics.1440}}.

\bibitem{ChoiCichocki2000}
S.~{Choi}, A.~{Cichocki}, Blind separation of nonstationary and temporally
  correlated sources from noisy mixtures, in: Neural Networks for Signal
  Processing X. Proceedings of the 2000 IEEE Signal Processing Society Workshop
  (Cat. No.00TH8501), Vol.~1, 2000, pp. 405--414.
\newblock \href {https://doi.org/10.1109/NNSP.2000.889432}
  {\path{doi:10.1109/NNSP.2000.889432}}.

\bibitem{ChoiCichocki2000b}
S.~{Choi}, A.~{Cichocki}, Blind separation of nonstationary sources in noisy
  mixtures, Electronics Letters 36~(9) (2000) 848--849.
\newblock \href {https://doi.org/10.1049/el:20000623}
  {\path{doi:10.1049/el:20000623}}.

\bibitem{ChoiCichockiBelouchrani2001}
S.~{Choi}, A.~{Cichocki}, A.~{Belouchrani}, Blind separation of second-order
  nonstationary and temporally colored sources, in: Proceedings of the 11th
  IEEE Signal Processing Workshop on Statistical Signal Processing (Cat.
  No.01TH8563), 2001, pp. 444--447.
\newblock \href {https://doi.org/10.1109/SSP.2001.955318}
  {\path{doi:10.1109/SSP.2001.955318}}.

\bibitem{Nordhausen2014}
K.~Nordhausen, On robustifying some second order blind source separation
  methods for nonstationary time series, Statistical Papers 55~(1) (2014)
  141--156.
\newblock \href {https://doi.org/https://doi.org/10.1007/s00362-012-0487-5}
  {\path{doi:https://doi.org/10.1007/s00362-012-0487-5}}.

\bibitem{TongLiuSoonHuang1991}
L.~{Tong}, R.~{Liu}, V.~C. {Soon}, Y.~{Huang}, Indeterminacy and
  identifiability of blind identification, IEEE Transactions on Circuits and
  Systems 38~(5) (1991) 499--509.
\newblock \href {https://doi.org/10.1109/31.76486}
  {\path{doi:10.1109/31.76486}}.

\bibitem{ErikssonKoivunen2004}
J.~{Eriksson}, V.~{Koivunen}, Identifiability, separability, and uniqueness of
  linear {ICA} models, IEEE Signal Processing Letters 11~(7) (2004) 601--604.
\newblock \href {https://doi.org/10.1109/LSP.2004.830118}
  {\path{doi:10.1109/LSP.2004.830118}}.

\bibitem{miettinen2015}
J.~Miettinen, S.~Taskinen, K.~Nordhausen, H.~Oja, Fourth moments and
  independent component analysis, Statistical Science 30~(3) (2015) 372--390.
\newblock \href {https://doi.org/10.1214/15-STS520}
  {\path{doi:10.1214/15-STS520}}.

\bibitem{IlmonenOjaSerfling2012}
P.~Ilmonen, H.~Oja, R.~Serfling, On invariant coordinate system {(ICS)}
  functionals, International Statistical Review 80~(1) (2012) 93--110.
\newblock \href {https://doi.org/10.1111/j.1751-5823.2011.00163.x}
  {\path{doi:10.1111/j.1751-5823.2011.00163.x}}.

\bibitem{muehlmann2020test}
C.~Muehlmann, F.~Bachoc, K.~Nordhausen, M.~Yi, Test of the latent dimension of
  a spatial blind source separation model (2020).
\newblock \href {http://arxiv.org/abs/2011.01711} {\path{arXiv:2011.01711}}.

\bibitem{CardosoSouloumiac1996}
J.-F. Cardoso, A.~Souloumiac, Jacobi angles for simultaneous diagonalization,
  SIAM Journal on Matrix Analysis and Applications 17~(1) (1996) 161--164.
\newblock \href {https://doi.org/10.1137/S0895479893259546}
  {\path{doi:10.1137/S0895479893259546}}.

\bibitem{IllnerMiettinenFuchsTaskinenNordhausenOjaTheis2015}
K.~Illner, J.~Miettinen, C.~Fuchs, S.~Taskinen, K.~Nordhausen, H.~Oja, F.~J.
  Theis, Model selection using limiting distributions of second-order blind
  source separation algorithms, Signal Processing 113 (2015) 95--103.
\newblock \href {https://doi.org/https://doi.org/10.1016/j.sigpro.2015.01.017}
  {\path{doi:https://doi.org/10.1016/j.sigpro.2015.01.017}}.

\bibitem{r_software}
{R Core Team}, \href{https://www.R-project.org/}{R: A Language and Environment
  for Statistical Computing}, R Foundation for Statistical Computing, Vienna,
  Austria (2019).
\newline\urlprefix\url{https://www.R-project.org/}

\bibitem{SpatialBSS_package}
C.~Muehlmann, K.~Nordhausen, J.~Virta,
  \href{https://CRAN.R-project.org/package=SpatialBSS}{SpatialBSS: Blind Source
  Separation for Multivariate Spatial Data}, {R} package version 0.9-0 (2020).
\newline\urlprefix\url{https://CRAN.R-project.org/package=SpatialBSS}

\bibitem{JADE_package}
J.~Miettinen, K.~Nordhausen, S.~Taskinen, Blind source separation based on
  joint diagonalization in {R}: The packages {JADE} and {BSSasymp}, Journal of
  Statistical Software 76~(2) (2017) 1--31.
\newblock \href {https://doi.org/10.18637/jss.v076.i02}
  {\path{doi:10.18637/jss.v076.i02}}.

\bibitem{RandomFields_package}
M.~Schlather, A.~Malinowski, P.~J. Menck, M.~Oesting, K.~Strokorb, Analysis,
  simulation and prediction of multivariate random fields with package
  {RandomFields}, Journal of Statistical Software 63~(8) (2015) 1--25.
\newblock \href {https://doi.org/10.18637/jss.v063.i08}
  {\path{doi:10.18637/jss.v063.i08}}.

\bibitem{Cardoso1989}
J.~{Cardoso}, Source separation using higher order moments, in: International
  Conference on Acoustics, Speech, and Signal Processing,, 1989, pp. 2109--2112
  vol.4.
\newblock \href {https://doi.org/10.1109/ICASSP.1989.266878}
  {\path{doi:10.1109/ICASSP.1989.266878}}.

\bibitem{NordhausenVirta2019}
K.~Nordhausen, J.~Virta, An overview of properties and extensions of {FOBI},
  Knowledge-Based Systems 173 (2019) 113--116.
\newblock \href {https://doi.org/https://doi.org/10.1016/j.knosys.2019.02.026}
  {\path{doi:https://doi.org/10.1016/j.knosys.2019.02.026}}.

\bibitem{IlmonenEtAl2010}
P.~Ilmonen, K.~Nordhausen, H.~Oja, E.~Ollila, A new performance index for
  {ICA}: Properties, computation and asymptotic analysis, in: V.~Vigneron,
  V.~Zarzoso, E.~Moreau, R.~Gribonval, E.~Vincent (Eds.), Latent Variable
  Analysis and Signal Separation, Springer, 2010, pp. 229--236.
\newblock \href {https://doi.org/https://doi.org/10.1007/978-3-642-15995-4_29}
  {\path{doi:https://doi.org/10.1007/978-3-642-15995-4_29}}.

\bibitem{LietzenVirtaNordhausenIlmonen2020}
N.~Lietzen, J.~Virta, K.~Nordhausen, P.~Ilmonen, Minimum distance index for
  {BSS}, generalization, interpretation and asymptotics, Austrian Journal of
  Statistics 49~(4) (2020) 57–68.
\newblock \href {https://doi.org/10.17713/ajs.v49i4.1130}
  {\path{doi:10.17713/ajs.v49i4.1130}}.

\bibitem{ReimannEtAl2014}
C.~Reimann, M.~Birke, A.~Demetriades, P.~Filzmoser, P.~O'Connor (Eds.),
  Chemistry of Europe's Agricultural Soils, Part A, Schweizerbart Science
  Publishers, 2014.

\bibitem{robCompositions_paper}
P.~Filzmoser, K.~Hron, M.~Templ, Applied Compositional Data Analysis. With
  Worked Examples in {R}, Springer, 2018.

\bibitem{muehlmann2020independent}
C.~Muehlmann, K.~Fačevicová, A.~Gardlo, H.~Janečková, K.~Nordhausen,
  Independent component analysis for compositional data (2020).
\newblock \href {http://arxiv.org/abs/2007.00456} {\path{arXiv:2007.00456}}.

\bibitem{comp_book}
J.~Aitchison, The Statistical Analysis of Compositional Data, Blackburn Press,
  2003.

\bibitem{PanMatilainenTaskinenNordhausen2021}
Y.~Pan, M.~Matilainen, S.~Taskinen, K.~Nordhausen, A review of second-order
  blind identification methods, WIREs Computational Statistics n/a (2021)
  e1550.
\newblock \href {https://doi.org/https://doi.org/10.1002/wics.1550}
  {\path{doi:https://doi.org/10.1002/wics.1550}}.

\bibitem{muehlmann2020cokriging}
C.~Muehlmann, K.~Nordhausen, M.~Yi, On cokriging, neural networks, and spatial
  blind source separation for multivariate spatial prediction, IEEE Geoscience
  and Remote Sensing Letters (2020).
\newblock \href {https://doi.org/10.1109/LGRS.2020.3011549}
  {\path{doi:10.1109/LGRS.2020.3011549}}.

\bibitem{PfisterWeichwaldBuhlmannScholkopf2019}
N.~Pfister, S.~Weichwald, P.~B{{\"u}}hlmann, B.~Sch{{\"o}}lkopf, Robustifying
  independent component analysis by adjusting for group-wise stationary noise,
  Journal of Machine Learning Research 20~(147) (2019) 1--50.
\newblock \href {https://doi.org/https://doi.org/10.3929/ethz-b-000374036}
  {\path{doi:https://doi.org/10.3929/ethz-b-000374036}}.

\end{thebibliography}


\end{document}